\documentclass[prd,aps,twocolumn,preprintnumbers,amsmath,amssymb]{revtex4}
\usepackage{graphicx,psfrag,hyperref}
\usepackage{dcolumn}
\usepackage{bm,bbm}
\usepackage{xcolor}
\begin{document}

\title{ The impact of electric currents on Majorana dark matter at freeze out }

\author{Lukas Karoly}
\author{David C. Latimer}

\affiliation{Department of Physics, University of Puget Sound,
Tacoma, WA 98416-1031
}

\newcommand*{\sech}{\mathop{\mathrm{sech}}\limits}
\newcommand*{\balpha}{\boldsymbol{\alpha}} 
\newcommand*{\dilog}{\mathrm{Li}_2}
\newcommand{\qslash}{\not{\hbox{\kern-3pt $q$}}}
\newcommand{\kslash}{\not{\hbox{\kern-3pt $k$}}}
\newcommand{\pslash}{\not{\hbox{\kern-2pt $p$}}}
\newcommand{\delslash}{\not{\hbox{\kern-3pt $\partial$}}}
\newcommand{\Dslash}{\not{\hbox{\kern-3pt $D$}}}
\newcommand{\gmn}{g^{\mu \nu}}
\newcommand{\Pslash}{\not{\hbox{\kern-2.3pt $P$}}}
\newcommand{\Kslash}{\not{\hbox{\kern-2.3pt $K$}}}
\newcommand{\Pslashsup}{^\not{\hbox{\kern-0.5pt $^P$}}}
\newcommand{\Poddup}{^\not{\hbox{\kern-0.5pt $^\mathcal{P}$}}}
\newcommand{\be}{\begin{equation}}
\newcommand{\ee}{\end{equation}}
\newcommand{\al}[1]{\begin{align}#1\end{align}}
\newcommand{\gv}{\gamma^5}
\newcommand{\bsig}{\boldsymbol{\sigma}}
\def\rcurs{{\mbox{$\resizebox{.16in}{.08in}{\includegraphics{ScriptR}}$}}}
\def\brcurs{{\mbox{$\resizebox{.16in}{.08in}{\includegraphics{BoldR}}$}}}
\def\hrcurs{{\mbox{$\hat \brcurs$}}}
\newcommand{\brp}{\mathbf{r}_+}
\newcommand{\brm}{\mathbf{r}_-}
\newcommand{\diff}{\mathrm{d}}
\newcommand{\bs}{\bigskip}
\newcommand{\beps}{\boldsymbol{\epsilon}}

\begin{abstract}
Thermal relics with masses in the GeV to TeV range remain  possible candidates for the Universe's dark matter (DM).  These neutral particles are often assumed to have vanishing electric and magnetic dipole moments so that they do not interact with single real photons, but the anapole moment can still be nonzero, permitting interactions with single virtual photons.  This anapole moment allows for p-wave annihilation of  DM into standard model particles, and the DM  interacts with external electric currents via the anapole moment.
Moving beyond their static electromagnetic properties, these particles generically have non-zero polarizabilities which mediate interactions with two real photons; in particular, spin-dependent polarizibilities admit s-wave annihilation of the  DM into two photons.  When the Universe cools from a temperature on the order of the DM mass to freeze out, the DM is in thermal equilibrium with the background plasma of particles, but the comoving DM density decreases due to annihilation.  If a collection of initially unpolarized DM particles were subjected to an electric current, then the DM medium would become partially polarized, according to the Boltzmann distribution, with a slight excess of anapole moments aligned with the current, relative to those anti-aligned.  
For this region of partially polarized DM particles, the s-wave annihilation mode becomes partially suppressed because it requires a state of vanishing angular momentum.  As a consequence, the decreased DM annihilation rate in this region will result in an excess of DM density, relative to an unpolarized region, as  DM drops out of thermal equilibrium. We explored this relative change of DM density for DM that is subjected to an electric current through freeze out.

  \end{abstract}

\maketitle

\section{Introduction}

A concordance of observations point to a universe whose matter content is overwhelmingly comprised of some new type of particles, outside the standard model \cite{pdg2022}.  The constraints on this new matter are few:  it must be non-relativistic, stable, and relatively weakly interacting.  A recent history of dark matter, including a discussion of various dark matter (DM) models, can be found in Ref.~\cite{dm_history}.
Early on, one class of DM models, weakly interacting massive particles (WIMPs), was theoretically well motivated, in part because of its potential tie in with supersymmetry \cite{susy_dm}.  Additional motivation for WIMPs  stemmed from a coincidence often called the ``WIMP miracle."  If DM were a thermal relic, then weak-scale masses and annihilation cross sections for the DM candidate naturally result in the correct relic density of DM observed today.  This coincidence is compelling, but we note that a much wider class of WIMPless models can  satisfy  the same relic density constraint, e.g. Ref.~\cite{wimpless}.

Despite a multimodal approach to DM detection, no definitive DM signal yet exists, though some observations show tantalizing hints. 
Perhaps because of  theoretical prejudice, many direct DM detection experiments focus upon the WIMP parameter space, broadly construed to include DM masses between the GeV scale up to a few TeV.   Aside from a few exceptions \cite{dama, dama2, cogent, cogent2, cdms_si}, decades of direct detection experiments have found no evidence for DM, and as a result, strict limits on the DM-nucleus interaction cross section \cite{lux, xenon_2018, pandax_4t} have ruled out many WIMP DM models.  In addition to direct detection experiments, particle colliders also place stringent constraints on DM models. In particular, the LHC has produced no particles outside the standard model, making tenuous  the notion that DM is comprised by supersymmetry's neutralino \cite{susy_lhc}.    
One other tack to assess the presence of dark matter is via indirect detection experiments in which telescopes search for high energy cosmic rays or photons. If signals  cannot be attributable to standard astrophysical sources, then they may be due to DM annihilation.  In some instances, this results on constraints  on the DM annihilation cross section, as with the observations of dwarf spheroidal galaxies from the Fermi Large Area Telescope (Fermi-LAT) for DM masses below 100 GeV \cite{fermiLAT}, while Fermi-LAT observations of the Galactic Center hint at the possibility of a DM annihilation \cite{hooper}.  Additionally, observations of antiprotons in the AMS-02 detector \cite{ams02}  could also signal DM annihilation for a DM mass around 80 GeV \cite{cuoco}.

Because of the severe constraints imposed by direct detection experiments and particle colliders, axion dark matter models are, perhaps, eclipsing WIMP models in terms of their favorability \cite{axions}, but the WIMP paradigm is not entirely dead because there is still viable parameter space remaining \cite{roszkowski, arcadi, blanco}.   With a narrowing parameter space, present-day modelers are opting to explore the WIMP paradigm with either simplified models or from the perspective of effective field theory (EFT) in which the modelers remain agnostic to a particular UV completion of the theory \cite{desimone}. 

In an EFT analysis of DM, one couples DM directly to standard model (SM) particles at low enegies via, often, dimensionful effective couplings that depend on a high-energy scale $\Lambda$.  As long as interaction energies are well below $\Lambda$, the effective interactions faithfully capture the relevant physics.  For neutral dark matter, the leading order electromagnetic interactions in an EFT occur through their static electromagnetic properties. Electric and magnetic dipole moments proceed through mass dimension-5 operators, and several DM modelers  have explored the possibility that DM predominantly interacts through such moments \cite{raby1987,raby1988, pospelov, sigurdson2004, sigurdson2004err, masso2009, heo2010, fitzpatrick2010, barger2011, barger2012, fortin2012, weiner2013, delnobile2014,   kopp2014, gresham2014, matsumoto2014, antipin2015}.  If the DM candidate is a Majorana fermion, then both the electric and magnetic dipole moments must vanish identically, and its sole static electromagnetic property is its anapole moment \cite{bk84,nieves}.  Anapole interactions of DM have been studied in Refs.~\cite{pospelov, fitzpatrick2010, ho2013, delnobile2014, kopp2014, gresham2014, gao2014, matsumoto2014,  sandick2016, gelmini2017, alves, blanco}.  Because the  anapole interaction arises from a dimension-6 operator, the anapole moment is suppressed by $\Lambda^2$ which further suppresses these leading order interactions for Majorana fermion DM. 
Additionally, one can choose model parameters so that annihilation precedes primarily through p-wave modes \cite{ho2013}, and because its annihilation is velocity suppressed, it can more easily evade indirect detection constraints \cite{blanco}.

Moving beyond the static electromagnetic properties of fermions, higher order electromagnetic interactions with Majorana fermions are mediated by operators that are dimension-7 and beyond \cite{pospelov, weiner2013,desimone}. In particular, a fermion can interact with two real photons via  its polarizabilities.   There are six two-photon interactions, two spin independent and four spin dependent, that are separately invariant under charge conjugation ($\mathcal{C}$), parity ($\mathcal{P}$), and time reversal ($\mathcal{T}$) \cite{ragusa}, and there an additional ten more polarizability terms that are either $\mathcal{C}$-odd and $\mathcal{T}$-odd or $\mathcal{C}$-odd and $\mathcal{P}$-odd  \cite{gorchtein}.  The self-conjugate nature of the Majorana fermion does limit its interaction with two real photons somewhat, requiring the four  $\mathcal{C}$-odd and $\mathcal{P}$-odd polarizabilites to vanish, but that still leaves a dozen modes unconstrained \cite{maj_2photon}.  

In this paper, we will focus upon Majorana fermion DM with a non-zero anapole moment and non-zero polarizabilities, but we will restrict our considerations 
to the six polarizabilities that separately preserve $\mathcal{C}$, $\mathcal{P}$, and $\mathcal{T}$ symmetries.  Of these six, the two spin-independent polarizabilities arise from a dimension-7 interaction, $\mathcal{L}_\text{SI pol} \sim \frac{1}{\Lambda^3} \chi^\dagger \chi F^{\mu \nu}F_{\mu \nu}$, that gives rise to the low-energy interaction Hamiltonian, expressed in terms of the electric and magnetic fields, $H_\text{SI pol} \sim \alpha_E E^2 + \beta_M B^2$.  The spin-independent electric and magnetic polarizabilities, $\alpha_E$ and $\beta_M$, are order $\mathcal{O}\left(\frac{1}{\Lambda^3}\right)$.  The four spin-dependent polarizabilities arise from interactions that depend upon {\em derivatives} of the electric and magnetic fields. As a consequence, the four spin-dependent polarizabilities, $\gamma_j$, are nominally $\mathcal{O}\left(\frac{1}{\Lambda^4}\right)$, though their precise mass dependence depends on the particular UV completion of the theory \cite{maj_2photon}.  It is these spin-dependent polarizabilities that allow for  s-wave annihilation of two Majorana fermions into two photons.  Nominally, this s-wave mode is suppressed relative to p-wave annihilation, but depending upon the particular UV completion, s-wave annihilation can be comparable to the p-wave mode \cite{anapole_2photon}.

The coupling between an anapole moment and a real photon vanishes, so the leading order electromagnetic interaction for a Majorana fermion is via the exchange of a virtual photon with a charged particle.  Thus, at low energies,  Majorana fermions do not couple to electric or magnetic fields; they only couple to electric currents.  If a spin-$\frac{1}{2}$ Majorana fermion is immersed in a persistent electric current, there is a difference in the two spin states of the fermion, with the lower energy state corresponding to the one in which the anapole moment is aligned with the current.  In the presence of a background current,  a collection of Majorana fermions can be undergo a some level of polarization, at least in principle.

This polarization can only be achieved if there are mechanisms that allow the Majorana fermion to change spin states. For particles that are not in thermal equilibrium, such as DM after freeze out, only irreversible processes can allow the higher-energy anti-aligned anapole moments to flip spin to the lower energy aligned state.  Spontaneous  two-photon emission or single-photon emission via virtual Compton scattering are two such irreversible mechanisms; however, because the photon coupling occurs through the polarizabilities, the rates of these irreversible processes are extremely small \cite{walter}.  However, before freeze out when the DM is in equilibrium with the thermal bath, {\em reversible} spin-flip processes can lead to a partially polarized DM medium in the presence of a background current, assuming the spin-flip interactions happen at a sufficient rate.  The Boltzmann distribution would guarantee a slight excess of lower-energy states with spins aligned with the current.
 
Our focus in this paper will be on DM in the early Universe, specifically when the DM is in thermal equilibrium with the Universe from a temperature around the DM mass, $T\sim m_\chi$, until freeze out.  As the Universe cools and expands, whenever its temperature is around the DM mass, most SM particles no longer have sufficient energy to produce DM via annihilation.  As a consequence, the comoving DM density decreases as it continues to annihilate into SM particles.  This decrease in comoving DM density continues until DM drops out of thermal equilibrium with the rest of the Universe at freeze out. At this point, because DM annihilations become so rare, the comoving DM density becomes constant, yielding the relic density present today.   This relic density is largely determined by the DM annihilation cross section; a larger cross section results in a smaller relic density, and vice versa.  

In this paper, we explore the consequence that a persistent local background current can have on DM in this time before freeze out. If the DM polarization induced by the current is sufficiently large, then the s-wave mode of DM annihilation can be suppressed because the s-wave mode requires the annihilating particles to have opposite spin states.  Overall, the DM annihilation rate will be somewhat smaller in the presence of the current than otherwise, which would result in  a local relic density that is higher than regions without a background current.

\section{EFT interactions}

Anapole interactions arise from an effective Lagrangrian term $\mathcal{L}_\text{ana} = \frac{1}{2} \frac{g}{\Lambda^2} \bar{\chi}^\dagger \gamma^\mu \gamma^5 \partial^\nu F_{\mu \nu} $ \cite{zeldovich, bk84, nieves}.  The anapole moment is the dimension-2 coefficient $f_a = \frac{g}{\Lambda^2}$ that results from a UV complete theory in which the neutral fermion effectively couples to the photon field through, at least, a one-loop process.  In the UV-complete Lagrangian, there must be a parity-violating \cite{zeldovich} trilinear term that couples the Majorana fermion to a charged fermion and (vector or scalar) boson. In terms of the mass scale $\Lambda$ for the anapole moment, it is set by the dominant mass of the charged particle to which the Majorana fermion couples.  Examples of how this coefficient depends on the underlying UV physics can be found in Refs.~\cite{neutrino_EM_review,nulino_anapole,maj_2photon}.

At tree-level, Majorana fermions scatter charged particles via the exchange of a virtual photon.  At low energies, the resulting interaction Hamiltonian is $H_\text{ana} = -f_a \boldsymbol{\sigma} \cdot \mathbf{J}$, where $\boldsymbol{\sigma}$ are the Pauli spin matrices and $\mathbf{J}$ is the current density associated with the charged particle \cite{zeldovich, bk84, nieves}.  Given this, a background current establishes an energy difference $\mathcal{E} = 2 f_a J$ between the Majorana fermion states aligned and antialigned with the current.

Similarly, at tree level, Majorana fermion  annihilation into a charged SM particle-antiparticle pair proceeds through the coupling of the anapole moment to a virtual photon. This p-wave cross section has been computed previously in Refs.~\cite{ho2013,gao2014}, and we quote the results here.   The thermally averaged cross section is
\begin{equation}
\langle \sigma_p |v| \rangle = 16 \alpha N_\text{eff} f_a^2 m_\chi^2  \left(\frac{T}{m_\chi}\right). \label{pwave}  
\end{equation}
The  factor $N_\text{eff}$ accounts for all the kinematically available final states, weighted by the square of the particles' charges.  For $m_\chi < 80$ GeV, all final states are fermionic.  Annihilation into an electron-positron pair contributes $1$ to $N_\text{eff}$; annihilation into a quark-antiquark pair, whose charges are $\pm q e$, contributes $3 q^2$ to $N_\text{eff}$, where the factor of 3 accounts for color degrees of freedom.  For $m_\chi > 80$ GeV, we must include the possibility that the Majorana fermions can annihilate into $W$ bosons. We can accommodate this in $N_\text{eff}$ with the term $\frac{3}{4} m_\chi^2/m_W^2$ if we employ the approximation, as in Ref.~\cite{gao2014}, that $m_W \gg m_\chi$.

Moving beyond the anapole moment, we consider two photon interactions with a Majorana fermion. Spin-independent interactions arise from  the dimension-7 effective Lagrangian discussed above.  We are interested in the spin-dependent two-photon interactions because s-wave annihilation proceeds through these channels. The  spin-dependent terms that couple the Majorana fermion to two photons involve derivatives of the electromagnetic field, and they have been characterized in Ref.~\cite{prange} and elsewhere.  The four coefficients of these dimension-8 terms, the spin-dependent polarizabilities $\gamma_j$, carry mass dimension $[M]^{-4}$.  Ostensibly, these polarizabilities would seem to scale as $\Lambda^{-4}$, but in considering an explicit simplified UV complete theory, the reality is somewhat more nuanced.  As with the anapole moment, the effective two-photon coupling to a Majorana fermion arises from a direct coupling of this fermion to charged particles.  Let's suppose these charged particles have mass $\Lambda$ and $m$ with $\Lambda > m$.  From a simplified UV-completion \cite{anapole_2photon}, one finds that the polarizabilities can scale as $\sim \frac{1}{\Lambda^4}$ and $\sim \frac{1}{\Lambda^2m^2}$. 

These two mass scales in the spin-depenent polarizability coeffecients also make an appearance in the s-wave annihilation cross section.  Dimensional analysis suggests $\langle \sigma_s |v|\rangle \sim \gamma^2 m_\chi^6 \sim \tilde{g}^2\frac{m_\chi^6}{\Lambda^8}$ where  $\gamma$ is some linear combination of the polarizabilities and $\tilde{g}$ is a dimensionless coefficient.  From a UV complete theory, we note that there are scenarios in which the s-wave annihilation cross section is not as suppressed as one might naively assume: $\langle \sigma_s |v|\rangle \sim \tilde{g}^2\frac{m_\chi^6}{\Lambda^4 m^4}$ \cite{anapole_2photon}.  We will adopt the EFT s-wave annihilation rate to be 
\begin{equation}
\langle \sigma_s |v|\rangle = \tilde{g}^2\frac{m_\chi^2}{\Lambda^4 \mu^4}, \label{swave}
\end{equation}
where $\mu:= \frac{m}{m_\chi}$ with $1 < \mu < \frac{\Lambda}{m_\chi}$.  

We would like to compare the relative s- and p-wave annihilation rates in order to determine their impact upon the relic density for a thermal WIMP.  In the anapole interaction, the dimensionless coupling, $g$, must be small enough in order for perturbative calculations to be viable; we will set $g=1$.  For the polarizabilities, the corresponding dimensionless coupling, $\tilde{g}$, is relatable to $g$ in a UV-complete theory.  The polarizabilities arise from, at least, a four-vertex Feynman diagram while the anapole moment comes from a three-vertex diagram.  Given this, $\tilde{g}$ should involve an extra factor of $e$ relative to $g$ which would suppress $\tilde{g}$ by a factor $\mathcal{O}(10^{-1})$ relative to $g$. But, at the same time, $\tilde{g}$ may incorporate corrections that are logarithmic in the relevant mass scale that could be $\mathcal{O}(10)$ \cite{maj_2photon}.  Given this, we will also take $\tilde{g}=1$, admitting that a particular UV-complete theory might deviate from this value by a factor of 10.

With these couplings fixed, we find that s-wave annihilation into photons is larger than or comparable to p-wave annihilation into charged particles whenever
\begin{equation}
\mu^4 \lesssim \frac{1}{16} \frac{1}{\alpha N_\text{eff}} \frac{m_\chi}{T}.
\end{equation} 
To see what size $\mu$ makes the two annihilation channels comparable, we consider DM masses, $m_\chi$, between 5 GeV and 80 GeV because, in this mass range, $N_\text{eff}$ is  fixed at 6.67.  
As a figure of merit, thermal WIMPs typically fall out of thermal equilibrium in the early Universe for a temperature around $T_f \sim \frac{m_\chi}{20}$.  Given this, we see that s-wave annihilation can exceed the p-wave process for $m \le 2.3 m_\chi$.  The upshot is that the s-wave annihilation mode is subdominant unless $m$ is  $ \mathcal{O}(m_\chi)$.

\section{ Rate of reversible spin-flip processes  }

Before freeze out, DM is in thermal equilibrium with the rest of the universe, and if the DM medium were within a background external current, we would expect a slight polarization of the medium by virtue of the Boltzmann distribution.
But, if a DM medium is initially unpolarized and then subjected to a current, there must be a sufficient rate of spin-flip interactions to assist in establishing polarization. In particular, the spin-flip rate must be much larger than the Universe's expansion rate, $\Gamma_\text{spin flip} \gg H$.  In the radiation-dominated era, we have 
\begin{equation}
H = 1.66 g_\star^\frac{1}{2} \frac{T^2}{M_\text{Pl}},
\end{equation}
where $g_*$ represents the relativistic degrees of freedom at temperature $T$ and $M_\text{Pl}$ is the Planck mass \cite{kolb_turner}.

The anapole moment is spin dependent, so interaction, via a single virtual photon, with a charged SM particle in the relativistic plasma can change a Majorana fermion's spin orientation.  We first focus upon the interaction between the Majorana fermion $\chi$ and one species of relativistic fermion $\psi$ with charge $qe$.  We assume the background current is $\mathbf{J} = J \hat{\mathbf{z}}$, and we average over the initial spin of the charged fermion and sum over its final spin states.  
Below, we compute the amplitude for the process: $\chi(p,\downarrow)+ \psi(k) \to \chi(p',\uparrow) + \psi(k')$.  To do so, we make several simplifying assumptions.  In particular, we neglect the mass of the charged fermion because it is relativistic. Also, the Majorana fermion is non-relativistic which implies $|\mathbf{p}|, |\mathbf{p}'| \ll m_\chi$. Implementing these approximations, the leading order contribution to the squared amplitude for this process is
\begin{equation}
| \mathcal{M}_\psi|^2 \approx 4q^2 e^2 f_a^2 m_\chi^2 (E_\psi E_\psi' - k_z k_z' ).
\end{equation}
Integrating over the phase space for the final states, the total DM spin-flip cross section is
\begin{equation}
\sigma_{\psi \text{ flip}} = q^2 \alpha  f_a^2 E_\psi^2,  
\end{equation}
where $\alpha$ is the fine structure constant. 

The rate at which the charged SM fermion can flip the spin of Majorana fermion depends on the total interaction cross section as well as the flux of incident charged particles, $\Gamma_{\psi \text{ flip}} = n_\psi |v| \sigma_{\psi \text{ flip}}$.  At temperature $T$, the fermion number density in the radiation-dominated early Universe is given by 
\begin{align}
n_\text{fermion} =& \frac{3}{4} \frac{\zeta(3)}{\pi^2} g_\text{dof} T^3, \label{nfermion}  
\end{align}
where $\zeta$ is the Riemann zeta function and $g_\text{dof}$ represents the degrees of freedom for the particular  species   \cite{kolb_turner}. 
In the plasma, these fermions are incident upon the Majorana fermion from all directions with a thermal distribution of momenta, so we average over all possible $\mathbf{k}$ in the distribution.  The thermally averaged cross section becomes
\begin{equation}
 \langle  \sigma_{\psi \text{ flip}} \, |v|\rangle =  \frac{15  \zeta(5)}{\zeta(3)} q^2 \alpha   f_a^2T^2.
\end{equation}
With this averaged cross section, we now have the thermally averaged spin-flip rate, $\langle \Gamma_{\psi \text{ flip}} \rangle$, due to interaction with a single species of fermion in the early Universe.  At a given temperature, there could be a host of different relativistic fermion species.  We incorporate this through an incoherent sum of spin-flip rates induced by each individual relativistic species appropriately weighted by its charge, $q$.

For temperatures above 80 GeV, the $W$ boson is relativistic, so it can appreciably contribute to the spin-flip rate of the Majorana fermion. As with the charged fermion, the interaction between the $W$ boson and Majorana fermion involves the exchange of a virtual photon, $\chi(p,\downarrow)+ W(k) \to \chi(p',\uparrow) + W(k')$.  When computing the cross section for this proces, we make several simplifying assumptions.  As above, we keep only leading order terms for the fermion spin line, taking $|\mathbf{p}|, |\mathbf{p}'| \ll m_\chi$, and we assume the $W$ boson to be highly relativistic with $|\mathbf{k}|, |\mathbf{k}'| \gg m_W$.  For these relativistic bosons, the longitudinal polarization state will dominate with $\varepsilon_L^\mu \to \frac{k^\mu}{m_W} $. Implementing these assumptions, the leading contribution to the squared amplitude is
\begin{equation}
|\mathcal{M}_W|^2 \approx 4 e^2 f_a^2\frac{m_\chi^2}{m_W^4} (k \cdot k')^2 |\mathbf{S} \cdot (\mathbf{k} + \mathbf{k}')|^2,
\end{equation}
where we define the three-vector  $\mathbf{S} := (1, -i, 0)$
  Integrating over the phase space of the final states, we find the total cross section to be 
\begin{equation}
 \sigma_{W \text{ flip}}\, |v| = \frac{1}{6} \alpha f_a^2 \frac{E_W^4}{m_W^2} (5 - \cos 2 \theta) ,
\end{equation}
where $\theta$ is the polar angle of the initial boson's momentum, $\mathbf{k}$.  
When relativistic, the boson number density in the radiation-dominated early Universe is given by  \cite{kolb_turner}
\begin{align}
n_\text{boson} = &  \frac{\zeta(3)}{\pi^2} g_\text{dof} T^3 \label{nboson}.
\end{align} 
Averaging over the thermal distribution of relativistic $W$ bosons we find 
\begin{equation}
 \langle  \sigma_{W \text{ flip}}\, |v|\rangle = 300 \alpha f_a^2 \frac{\zeta(7)}{\zeta(3)} \frac{T^4}{m_W^2}.
 \end{equation}
For temperatures beyond $m_W$, we add to the thermally averaged fermion spin-flip rate the contributions from the interactions with the $W$ boson, $\langle \Gamma_{W \text{ flip}} \rangle = n_\text{boson}  \langle  \sigma_{W \text{ flip}}\, |v|\rangle $.  

\begin{figure}[h]
\includegraphics[width=8.6 cm]{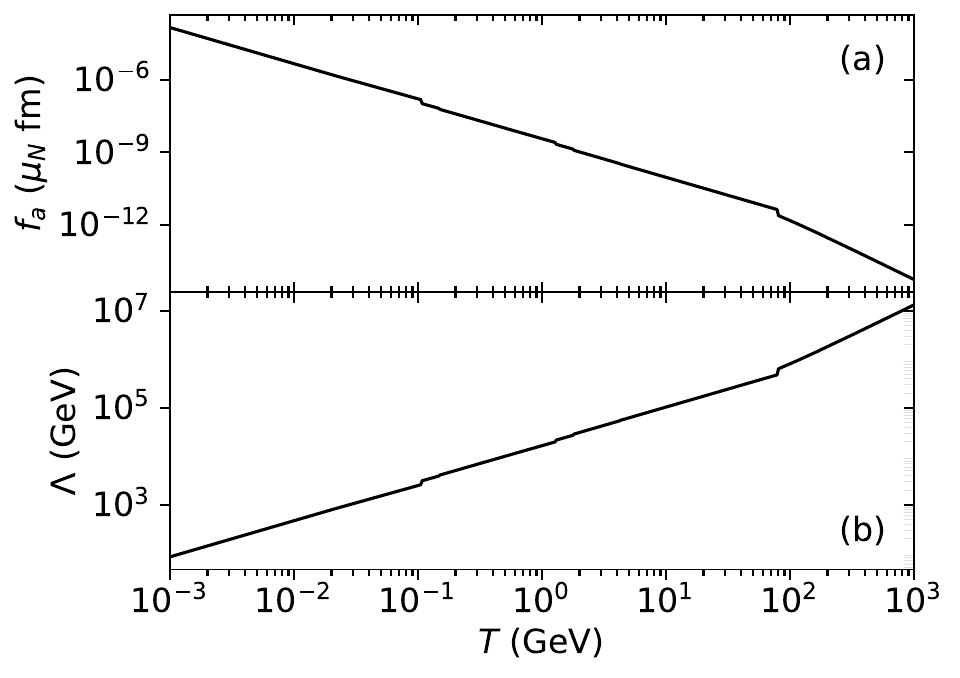}
\caption{(a) In the top panel, the curve plots the anapole moment at which the spin-flip rate of a non-relativistic Majorana fermion equals that of the Hubble parameter for a given temperature in the radiation-dominated era of the early Universe. (b) Setting $f_a = \frac{g}{\Lambda^2}$ with $g=1$, the curve in the lower panel shows the energy scale $\Lambda$ at which the spin-flip rate of a non-relativistic Majorana fermion equals that of the Hubble parameter for a given temperature in the radiation-dominated era of the early Universe. } \label{fig1}
\end{figure}

We now consider  the rate at which all relativistic particles can flip the spin of a non-relativistic Majorana fermion in the early Universe.  In comparing the spin-flip rate to the Universe's expansion rate, we find $\frac{\langle \Gamma_{\text{ flip}} \rangle }{H} \approx \alpha f_a^2 M_\text{Pl} T^3$ whenever only charged fermions are relativistic, $T < m_W$.  Beyond that temperature, we have $\frac{\langle \Gamma_{\text{ flip}} \rangle }{H} \approx \alpha f_a^2 \frac{M_\text{Pl}}{m_W^2} T^5$, accurate to within a factor of a few.  Using the full expression for the spin-flip rate, we plot in Fig.~\ref{fig1}(a) the anapole moment at which the spin-flip rate equals the Hubble parameter as a function of temperature. For anapole moments above this curve, spin-flip interactions are sufficiently frequent to allow a collection of Majorana fermions to partially polarize in a background current via thermalization.  

Treating the anapole moment as an effective interaction, $f_a = \frac{g}{\Lambda^2}$, we can constrain the energy scale for UV completion.  Setting $g=1$, we plot in Fig.~\ref{fig1}(b) the energy $\Lambda$ for which the Majorana fermion's spin-flip rate in the early Universe is equal to the Hubble expansion rate.  From the figure, we may determine, for a given temperature, the upper limit on $\Lambda$ for which  spin-flip interactions are sufficient to achieve Majorana fermion polarization.  This energy scale can also be used to inform our knowledge of additional, higher-order, effective electromagnetic interactions with the Majorana fermion.  In particular, it sets the scale for the DM's polarizabilities.

Considering a fixed $\Lambda$, then Fig.~\ref{fig1}(b) shows the {\em lowest} temperature at which the spin-flip and expansion rates are equal because $\frac{\langle \Gamma_{\text{ flip}} \rangle }{H} \sim T^3$ (or $\frac{\langle \Gamma_{\text{ flip}} \rangle }{H} \sim T^5$ for higher temperatures, $T> m_W$).   In what follows, we would like this spin-flip rate to be sufficiently large for temperatures through freeze out, so that polarization can be achieved up until the point at which dark matter decouples from the background thermal bath.  If we are to interpret the temperature in Fig.~\ref{fig1}(b) as the freeze-out temperature for a given dark matter candidate, then the constraints on $\Lambda$ will be more stringent.  We estimate these more stringent constraints below.

The freeze-out temperature, $T_f$, for a DM candidate is determined primarily by the thermally averaged annihilation cross section $\langle \sigma_\text{ann} |v|\rangle$. The cross section can be expanded in a power series for velocity because  DM is non-relativistic when it decouples from the thermal background.  Here we assume one velocity mode dominates the annihilation cross section and  parametrize the cross section in terms of the background temperature by virtue of $v \sim T^\frac{1}{2}$, viz., $  \langle \sigma_\text{ann} |v|\rangle = \sigma_0 x^{-n}$  where $x = \frac{m_\chi}{T}$.  If s-wave annihilation dominates, then $n=0$; for p-wave, $n=1$; and so on.

To precisely determine freeze out and the relic DM density, one must solve the Boltzmann equation, as discussed  below in Sec.~\ref{boltz_sec}.  However, estimates, accurate to a few percent, do exist.  In particular, the freeze-out temperature and relic number density $Y= n/s$ (relative to the entropy density $s$) are given by
\begin{align}
x_f & \approx  \log[ (n+1)  a \lambda ] -
\left(n+\tfrac{1}{2}\right)\log[ \log[ (n+1)  a \lambda] ]  \label{xf} \\
Y_\infty & \approx  \frac{(n+1)} {\lambda} x_f^{n+1} \label{Yf}
\end{align}
where $a = 0.289\, g_*^{-1}$ and $\lambda =  0.264\, g_*^{1/2}M_\text{Pl} m_\chi \sigma_0$ \cite{Scherrer:1985zt, kolb_turner}.

\begin{figure}[h]
\includegraphics[width=8.6 cm]{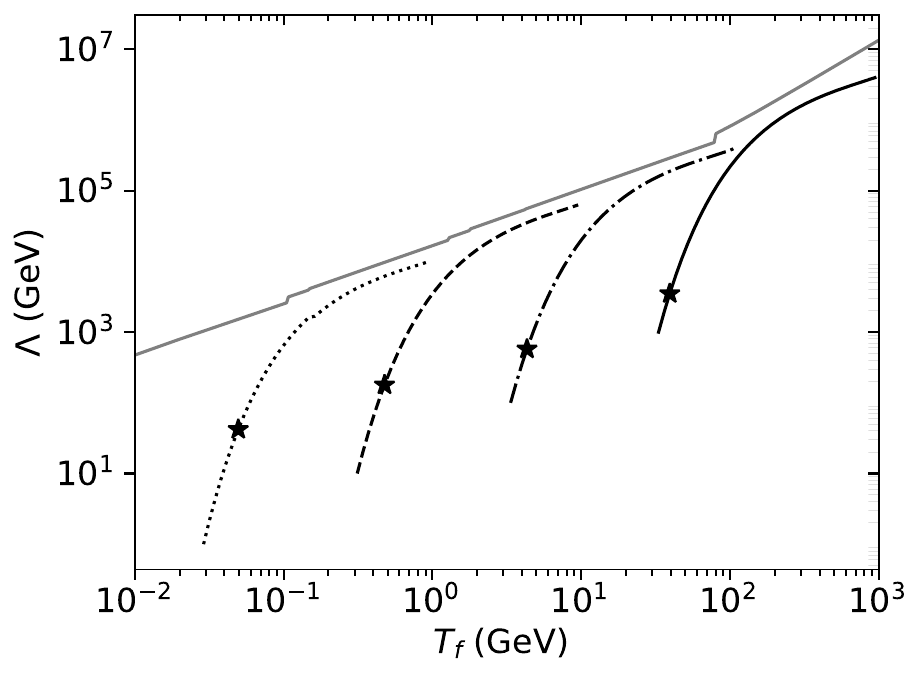}
\caption{For a given DM mass $m_\chi$, the black curves show the value of $\Lambda$ that yields a given freeze-out temperature, $T_f$. For the dotted curve, the mass is 1 GeV; for the dashed curve, the mass is 10 GeV; for the dot-dashed curve, the mass is 100 GeV; for the solid (black) curve, the mass is 1 TeV.  The value of $\Lambda$ that reproduces the relic DM density for a given mass is denoted by $\bigstar$. The solid gray curve represents the spin-flip constraints on $\Lambda$ for a given temperature, reproduced from Fig.~\ref{fig1}(b) \label{fig2} }
\end{figure}

 For the models under consideration herein, we assume the p-wave contribution to the cross section to dominate, so we set $n=1$ and use the cross section in Eq.~(\ref{pwave}).  Given a particular  DM mass $m_\chi$, we can determine what value of energy scale $\Lambda$ yields a given freeze-out temperature $T_f > m_\chi$.  Figure \ref{fig2} contains these results for a range of DM masses from 1 GeV to 1 TeV.  We superpose on this plot the curve from Fig.~\ref{fig1}(b) which shows the upper limit for $\Lambda$ at which the spin-flip rate is sufficient to achieve thermalization. Not surprisingly, for the models under consideration, the spin-flip rate is sufficiently large through the freeze-out temperature.   Additionally, we consider the values of $\Lambda$ and $m_\chi$ that reproduce the relic DM mass density present today $\rho_\text{DM} = \Omega_\text{DM} \rho_\text{crit}$, where  $\Omega_\text{DM}$ is the DM fraction of the energy budget and $\rho_\text{crit}$ is the critical energy density \cite{pdg2022}.  If we assume our DM candidate is to reproduce the relic DM density, then the energy scale $\Lambda$ is sufficiently small so that the DM candidate interacts  through freeze out to thermalize in a background current.

\section{DM density in a current \label{boltz_sec}}

We must use the Boltzmann equation to  precisely determine the evolution of the DM density from the time it becomes non-relativistic through freeze out.
As above, we express as $Y = n/s$ the DM number density relative to the entropy density, $s$. Given this, the first moment of the Boltzmann equation can be written as 
\begin{equation}
\frac{\diff}{\diff x} Y = -  \left( 0.602g_*^{-\frac{1}{2}}\frac{M_\text{Pl}}{m_\chi^2} \right) \langle \sigma_\text{ann} |v| \rangle s x (Y^2 - Y_\text{eq}^2).  \label{Yeqn1}
\end{equation}
The term $Y_\text{eq}(x)$ tracks the equilibrium number density which we take as $Y_\text{eq} = a\, x^\frac{3}{2} e^{-x}$ in the non-relativistic regime ($x\gg 3$), where again $a = 0.289 g_*^{-1}$.
In what follows, we would like to determine the impact of a locally polarized region of DM on the local relic DM density, but the  
the Boltzmann equation is derived under the assumptions of homogeneity and isometry. The presence of a local current clearly violates that, but because the inhomogeneities we introduce are so small,  we will treat the  current as a local perturbative term in the Boltzmann equation.

Though the p-wave annihilation cross section typically dominates, Eq.~(\ref{pwave}), we must also consider the s-wave mode, Eq.~(\ref{swave}).  For what follows, it will be useful to factor the cross section as
\begin{align}
\langle \sigma_\text{ann} |v| \rangle = & \sigma_0 x^{-1}(1 + b x),
\end{align}
where $\sigma_0 = \langle \sigma_\text{p} |v| \rangle  x$ and $b =  \langle \sigma_\text{s} |v| \rangle/\sigma_0$.  For the DM interactions considered herein, we find from Eq.~(\ref{pwave}) that $\sigma_0 = 16  \alpha m_\chi^2 N_\text{eff}/\Lambda^4$.   Then, from Eq.~(\ref{swave}), we compute $b =  \tilde{g}^2/(16  \alpha N_\text{eff} \mu^4)$.  If  the s-wave process is to be a subdominant correction through freeze out, then we must require $b x_f \ll 1$.  This  constrains the parameter $\mu$:   $\mu \gg  [x_f/(16 \alpha N_\text{eff})]^\frac{1}{4} \sim 2.6$.  Substituting the factored annihilation cross section in Eq.~(\ref{Yeqn1}), we have
 \begin{equation}
 \frac{\diff}{\diff x} Y = -  \lambda (1 + b x)   x^{-3}  (Y^2 - Y_\text{eq}^2),  \label{Yeqn2}
\end{equation}
where again $\lambda = \left( 0.264\, g_{*}^{\frac{1}{2}} M_\text{Pl} m_\chi \right)\sigma_0$.

\begin{figure}[h]
\includegraphics[width=8.6 cm]{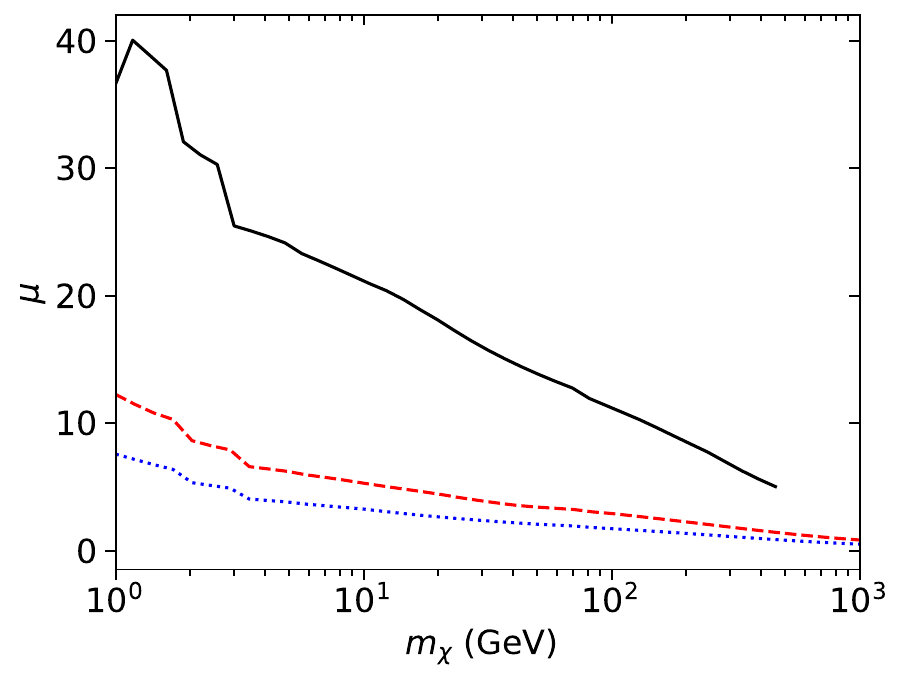}
\caption{ Bounds on $\mu$ derived from observational limits on the s-wave DM annihilation mode.  The solid (black) curve uses data from Ref.~\cite{ackermann}; the dashed (red) curve uses data from Ref.~\cite{damico}; and the dotted (blue) curve uses data from Ref.~\cite{slatyer}.   } \label{fig3}
\end{figure}

Aside from our desire to keep the s-wave annihilation mode sub-dominant, there are observational constraints on the annihilation of DM into two photons that come from a variety of sources.   In  search of  mono-energetic gamma rays from annihilating DM in the galactic halo, Fermi-LAT places the most stringent constraints on the annihilation into two photons; we use the most stringent constraints from the R3 region of interest in Ref.~\cite{ackermann}.
 Additionally, precise measurements of the cosmic microwave background (CMB) anisotropies from the Planck satellite  \cite{planck2015_cosmo} severely constrain the energy injection from DM annihilation at the time of recombination. An analysis of the CMB has resulted in stringent constraints on the s-wave DM annihilation cross section considered herein \cite{slatyer}.  
 Finally,  an absorption feature  has been observed in the 21-cm spectrum at high redshift;  this constrains DM s-wave annihilation because energy injection from annihilation would wash out the feature  \cite{damico}.  
 We use these constraints on the s-wave cross section to place lower bounds on the parameter $\mu$ in Eq.~(\ref{swave}).  To do so, we assume that, for a given DM mass $m_\chi$, the energy-scale $\Lambda$ is set by reproducing the DM relic density through p-wave annihilation only.  Then the limits from Fermi-LAT, the CMB, and 21-cm data bound $\mu$ as show in Fig.~\ref{fig3}

In the presence of a local current density, we need to modify the s-wave contribution to the cross section because it requires a spin-zero initial state and a current can partially polarize the DM medium before freeze out. In particular, suppose a current $J$ exists in a region. In thermal equilibrium, the number density of DM particles with spins aligned with the current, $n_\uparrow$, will exceed those anti-aligned, $n_\downarrow$, by an amount $n_\uparrow/n_\downarrow = \exp[\mathcal{E}/T] \approx 1 + \mathcal{E}/T$ where $\mathcal{E} = 2 f_a J$ is the energy difference between the aligned and anti-aligned states. The fractional relative difference in the two spin states is $\epsilon:= (n_\uparrow - n_\downarrow)/n \approx \mathcal{E}/(2T)$.  Then, in the Boltzmann equation, Eq.~(\ref{Yeqn2}), the s-wave annihilation in the presence of a current is suppressed by a factor of $(1-\epsilon)$
 \begin{equation}
 \frac{\diff}{\diff x} Y = -  \lambda [1 + b  (1- \epsilon) x]   x^{-3}  (Y^2 - Y_\text{eq}^2),  \label{Yeqn_current}
\end{equation}
where $\epsilon = f_a J/T$.

Before exploring the impact of a current in detail, it is worth considering the most extreme possibility in which  s-wave annihilation is prohibited by virtue of complete polarization of the DM medium; that is, we compare the $\epsilon =1 $ scenario (full polarization) with the $\epsilon =0$ scenario (no polarization). We will do this, first, using the estimates of $x_f$ and $Y_\infty$ from Eqs.~(\ref{xf}) and (\ref{Yf})  extended to include both s- and p-wave annihilation \cite{kolb_turner}
\begin{align}
x_f & \approx  \log[  2a \lambda ] -
\tfrac{3}{2}\log[ \log[ 2 a \lambda] ] +\log [1 + b \log[ 2a   \lambda ]]  \label{xf+} \\
Y_\infty & \approx  \frac{2} {\lambda} x_f^{2} \frac{1}{ (1 + 2 b x_f)  } \label{Yf+}
\end{align}
If the p-wave process dominates freeze out, then $x_f \sim \log[2 a \lambda]$, and if we continue with our previous approximation $b x_f \ll 1$, then the s-wave channel modifies the freeze-out temperature by $x_f \stackrel{\sim}{\mapsto}
x_f +b \log[ 2a   \lambda ]$.   Upon including the s-wave annihilation mode, the relic DM density should decrease by $Y_\infty \stackrel{\sim}{\mapsto} Y_\infty(1 - 2 b x_f)$.  It is most useful to cast these changes in terms of the fractional change in the relic density; full polarization of the DM medium (which turns off the s-wave mode) relative to the no-current scenario results in a fractional change of 
\begin{equation}
\frac{Y_\infty^{\epsilon=1} -Y_\infty^{\epsilon=0}}{Y_\infty} \approx  2 b x_f.\label{deltaYest}
\end{equation} 

Using some exemplar parameters, we would like to  numerically integrate the Boltzmann equation, Eq.~(\ref{Yeqn_current}), to confirm the accuracy of the estimates in Eqs.~(\ref{xf+} - \ref{deltaYest}).
Because Eq.~(\ref{Yeqn_current}) is an extremely stiff differential equation, it is easier to set $W = \log Y$ and instead integrate the equation \cite{steigman2012}
\begin{equation}
 \frac{\diff}{\diff x} W =  \lambda [1 + b(1-\epsilon) x]   x^{-3}  \left( e^{(2W_\text{eq} - W)} -e^W\right).  \label{Weqn}
\end{equation}
For parameters, we set $m_\chi = 100$ GeV. Annihilation dominated by the p-wave process reproduces the observed DM relic density for $\Lambda = 579$ GeV with a freeze-out temperature around $x_f \sim 23.1$.   If we use the most stringent constraints on $\mu$ derived from the Fermi LAT data \cite{ackermann}, then $\mu = 11.2$. From the estimate in Eq.~(\ref{deltaYest}), we expect a fractional increase in the relic DM density of  0.4\% for a fully polarized DM medium.
  Numerical calculations produce the same order of magnitude change.  Computing $Y(x = 1000)$ for both scenarios, we find that the p-wave only (full polarization) relic density is a factor of 0.1\% larger than  when both p- and s-wave annihilations (no current scenario) are considered.  (We note that the estimate of $Y_\infty$ from Eq.~(\ref{Yf+}) relative to the numerical computation of $Y(x=1000)$  differs by 3.5\%; however, the {\em relative} fractional change in the estimates using Eqs.~(\ref{Yf}, \ref{Yf+}) for the p-wave only and p- and s-wave computations for $Y_\infty$ yield the correct order of magnitude result.)

We now discuss  the impact of a local current upon the local relic DM density.  In order to do so, we must supply some details about the form of the current. We treat the current classically, assuming that it can be represented as the net drift of the relativistic charged species in the plasma, $J = e n_q v_\text{drift}$.  
The factor $n_q$ is the sum over the number density of all charged relativistic species, Eqs.~(\ref{nfermion}) and (\ref{nboson}), in the plasma at temperature $T$ weighted by their absolute charge.  We suppose that the current exists from $x = 1$ ($T = m_\chi$) through freeze out, $x\sim 20 - 25$, and we assume that the drift velocity suffers a redshift due to expansion so that $v_\text{drift}(x) = \frac{1}{x} v_\text{drift}(1)$.  With these assumptions, the current scales like $J \sim x^{-4}$, and overall, the perturbing term in Eq.~(\ref{Yeqn_current}) scales with $x$ as $\epsilon =  f_a J/T  \sim x^{-3}$.  

Before we  integrate Eq.~(\ref{Yeqn_current}) with the classical current, we will develop some approximations that allow us to estimate the relative change in the local relic density.  Following the arguments in Ref.~\cite{kolb_turner} that yield Eqs.~(\ref{xf+}) and (\ref{Yf+}), we can achieve the order of magnitude estimates
\begin{align}
x_f^\epsilon \approx &  x_f^{\epsilon=0} - b\epsilon  \log[ 2a   \lambda] \\
Y^\epsilon(x_f) \approx & Y^{\epsilon=0}(x_f) \left[    1 +\frac{1}{2} b\epsilon x_f    \right] 
\end{align}
where  $\epsilon$ is evaluated at $x_f^{\epsilon=0}$.   The fractional change in the local relic density in the presence of the assumed classical current is 
\begin{equation}
\frac{Y^{\epsilon}(x_f) -Y^{\epsilon=0}(x_f)}{Y(x_f)} \approx  \frac{1}{2} b\epsilon x_f.\label{deltaYest_current}
\end{equation} 

We would like to compare this crude estimate with a more robust solution of Eq.~(\ref{Yeqn_current}). The current introduces a small perturbation in the DM density $\delta Y = Y^{\epsilon} -  Y^{\epsilon=0}$, and it is sufficient to 
 linearize Eq.~(\ref{Yeqn_current}) with respect $\delta Y$, neglecting small quantities
\begin{equation}
 \frac{\diff}{\diff x} \delta Y = -  2 \lambda [1 + b  x]   x^{-3}   Y \delta Y  +  \epsilon \lambda   b x^{-2}  (Y^2 - Y_\text{eq}^2), \label{deltaY}
 \end{equation}
 where $Y = Y^{\epsilon=0}$ is the no-current DM density. 
 This first order non-homogenous ordinary differential equation can be solved with an integrating factor.
 Defining the functions \begin{align}
p(x) =& 2\lambda [1 + b x ]   x^{-3}   Y \label{px} \\
q(x) = & \epsilon \lambda b x^{-2}  (Y^2  - Y_\text{eq}^2), \label{qx}
\end{align}
then the solution to Eq.~(\ref{deltaY}) is
\begin{equation}
\delta Y(x) =  \int_{1}^x P(x)^{-1}  P(s) q(s) \mathrm{d} s \label{deltaY_ODE}
\end{equation}
where
\begin{equation}
P(x) = \exp\left[\int^x_{1} p(s)\mathrm{d}s\right].
\end{equation}
Examining Eq.~(\ref{deltaY_ODE}), we see that the approximation for $\delta Y$ is explicitly linear in $v_0$ and manifestly positive because $q(x)$ is positive.

To perform the integral in Eq.~(\ref{deltaY_ODE}), we first consider the factor
\begin{equation}
P(x)^{-1} P(s) = \exp \left[ -\int_s^x p(t) \diff t\right]. \label{pinvp}
\end{equation}
For the parameters under consideration $p(x)$ is extremely large, ranging from $\sim 10^{18}$ around $x=1$ to $\sim 10^8$ around $x = x_f$ for the DM mass of 100 GeV (and associated parameters considered above).  Because $p(x)$ is so large, the factor $P(x)^{-1} P(s)$ vanishes, for all practical purposes, except at $s = x$, where $P(x)^{-1}P(s=x) = 1$.  Because of this feature, only the value $q(x)$ is of any consequence in the integrand
\begin{equation}
\delta Y(x)  \approx q(x)  \int_{1}^x P(x)^{-1}  P(s)  \mathrm{d} s \label{deltaY_ODE2}.
\end{equation}
The remaining integral in Eq.~(\ref{deltaY_ODE2}) can be accurately estimated by a Taylor expansion of the argument of the exponential  in Eq.~(\ref{pinvp}) about $s =x$
\begin{equation}
P(x)^{-1} P(s) \approx \exp \left[ (s-x) p(x)  \right]. 
\end{equation}
We can then estimate the integral
\begin{equation}
\int_{1}^x P(x)^{-1}  P(s)  \mathrm{d} s \approx \frac{1}{p(x)}
\end{equation}
for $x \gg 1$.
This yields an estimate for $\delta Y$ of
\begin{equation}
\delta Y (x) = \frac{q(x)}{p(x)}
\end{equation}
for $x\gg 1$.  
Using the definitions of $p$ and $q$ in Eqs.~(\ref{px}, \ref{qx}), we find at freeze out
\begin{equation}
\delta Y (x_f) = \frac{\epsilon  b x_f  [Y^2(x_f)  - Y^2_\text{eq}(x_f)] }{2 [1 + b x_f ]      Y(x_f)},  
\end{equation}
where $\epsilon$ is evaluated at $x_f$.  For small $b$, we can  approximate this as
\begin{equation}
\frac{\delta Y (x_f)}{Y(x_f)} = \frac{1}{2} \epsilon  b x_f  \left[1 - \frac{Y^2_\text{eq}(x_f)}{Y^2(x_f)}\right],   \label{delYsoln}
\end{equation}
In Eq.~(\ref{delYsoln}), if we further neglect $Y_\text{eq}^2$ relative to $Y^2$ at freeze out, then we recover our crude estimate from Eq.~(\ref{deltaYest_current}): $\dfrac{\delta Y(x_f)}{Y(x_f)}\approx \frac{1}{2} \epsilon  b x_f$.  Executing the calculation for $\delta Y/Y$ in Eq.~(\ref{delYsoln}) for DM mass $m_\chi = 100$ GeV (and associated parameters considered above), we find that the result is about $0.58$ times the crude estimate, $\frac{1}{2} \epsilon  b x_f$. 
\begin{figure}[h]

\includegraphics[width=8.6 cm]{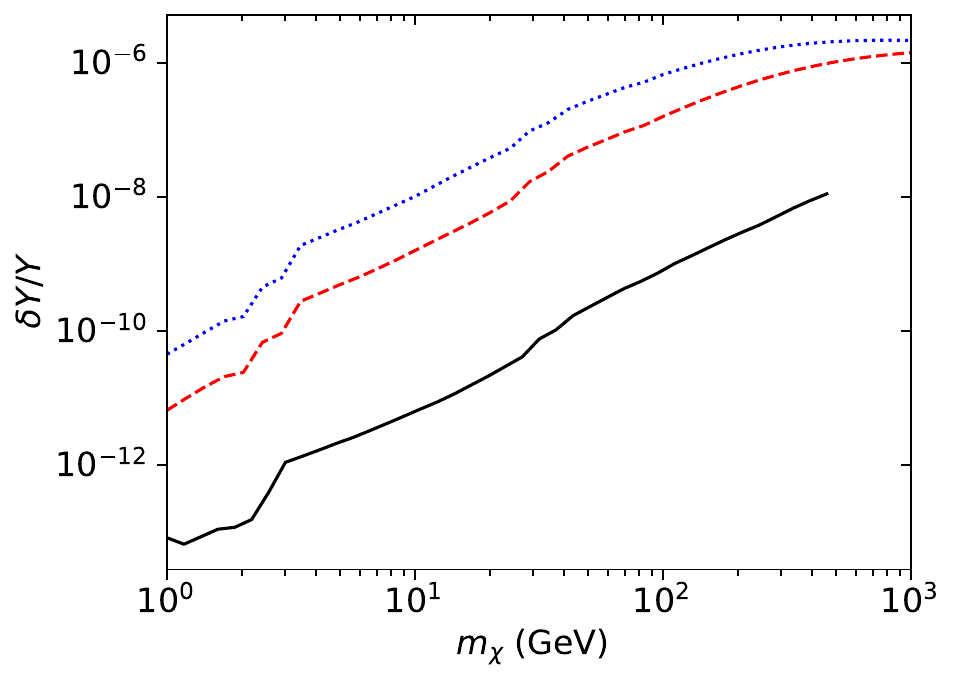}
\caption{  The local relative change in DM density at freeze out due to the presence of a classical current with $v_\text{drift}(1) =1$. For a given $m_\chi$, the parameters are chosen to reproduce the observed relic DM density and satisfy constraints on s-wave annihilation derived from  Fermi-LAT data (the solid (black) curve) \cite{ackermann},  21-cm spectral data  (the dashed (red) curve) \cite{damico}, and CMB data (the dotted (blue) curve) \cite{slatyer}.   
} \label{fig4}
\end{figure}

With Eq.~(\ref{delYsoln}), we are now able to determine how a local  electric current can impact the local DM density through freeze out.   As noted previously, we model the current classically as a steady current from $x=1$ through freeze out, modulo a decreasing plasma density and red shifted drift velocity, whose initial value is $v_\text{drift}(1)$.  For our calculations, we take $v_\text{drift}(1)=1$.  Because 
$\delta Y/Y$ is manifestly linear in $v_\text{drift}(1)$, one can easily scale our results to accommodate more realistic values for the drift velocity.  With this assumption, we present our results in Fig.~\ref{fig4}.  For a given DM mass, we determine the mass scale $\Lambda$ by fixing the relic DM density to the observed value, assuming p-wave annihilations determine this. Then, we set the mass ratio $\mu$ to satisfy the constraints on s-wave annihilation derived from Fermi-LAT data \cite{ackermann},  CMB data from the Planck satellite  \cite{planck2015_cosmo, slatyer}, and observations of the 21-cm line \cite{damico}.

\section{Discussion and conclusion}

Herein, we considered the impact that a local current might have on the evolution of the DM density around DM freeze out. The anapole moment of a Majorana fermion DM candidate tends to align with external currents provided there are sufficient spin-flip interactions to thermalize the DM. We find for the model parameters under consideration DM can thermalize, so that the DM states, described by a Boltzmann distribution, can lead  to a partially polarized DM medium. This partial polarization is of consequence for  the available DM annihilation channels.  

Rather generically, a Majorana fermion DM candidate can interact with real photons through higher order processes.  In particular, DM can annihilate into two real photons in an s-wave process by virtue of its spin-dependent polarizabilities.  In a partially polarized DM medium, this s-wave annihilation channel is somewhat suppressed because it requires the initial DM states to have opposite spins.  As a consequence, the overall DM annihilation rate is smaller than it would be if no current were present, and as the Universe cools and expands, the lower annihilation rate will result in a slightly higher relic DM density than in a no-current region.

Referring to Fig.~\ref{fig4}, we find that, for an initial charge carrier drift speed approaching the speed of light, the relative over density of DM in the current bearing region ranges from $\sim 10^{-13}$ to $\sim 10^{-6}$ depending on the DM mass and s-wave constraint used. Generally, the constraints on $\mu$ decrease with DM mass, and because $\delta Y \sim \mu^{-4}$, the size of $\delta Y$ increases substantially with $m_\chi$.   A density variation of order $10^{-6}$ is extremely large at this time in the early Universe, but this is based upon an unrealistic drift speed.  Our results scale linearly with initial drift speed, so it is trivial to determine the impact upon more realistic current magnitudes.

In terms of the assumptions in our calculations, we assume a current that persists from temperature $T =  m_\chi$ to freeze out $T \sim \frac{1}{20} m_\chi$. The density change $\delta Y$, however, is most impacted by the existence of the current around freeze out, so the assumed longevity of the current in our calculations could be relaxed without significantly impacting the results.  Additionally, we modeled our current as a simple classical current density; more complex models could also fit into our existing work.   Finally, throughout we assumed that the p-wave annihilation mode determines the relic density in our calculations. This is an excellent approximation when using the constraints on $\mu$ derived from the observations in Ref.~\cite{ackermann}, but for DM masses greater than a few hundred GeV, the approximation is strained if using the constraints on $\mu$ derived from Refs.~\cite{slatyer, damico}.

The premise of our entire work requires the presence of substantial electric currents in the early Universe around DM freeze out, but the actual existence of such  currents is beyond the scope of this work. We do find, in the literature, 
arguments supporting the existence of electric currents in the early Universe specifically during  the inflationary period \cite{PhysRevD.37.2743}, around the electroweak phase transition \cite{PhysRevD.55.4582,PhysRevD.53.662}, and around the QCD phase transition \cite{QCDtrans_currents, PhysRevD.50.2421,PhysRevD.55.4582,PhysRevD.77.043529}.  These currents should be short-lived, compared to the Hubble time \cite{siegel_fry}.  For our purposes, long-lived currents are not crucial to the results, rather the strength of the current and the time at which it occurs are the more important factors.

\section{ACKNOWLEDGMENTS}
DCL thanks the Kavli Institute for Theoretical Physics for its hospitality during the completion of this work. 
This research was supported in part by the National Science Foundation under Grant No. NSF PHY-1748958.

\bibliography{biblio}

\begin{thebibliography}{69}
\expandafter\ifx\csname natexlab\endcsname\relax\def\natexlab#1{#1}\fi
\expandafter\ifx\csname bibnamefont\endcsname\relax
  \def\bibnamefont#1{#1}\fi
\expandafter\ifx\csname bibfnamefont\endcsname\relax
  \def\bibfnamefont#1{#1}\fi
\expandafter\ifx\csname citenamefont\endcsname\relax
  \def\citenamefont#1{#1}\fi
\expandafter\ifx\csname url\endcsname\relax
  \def\url#1{\texttt{#1}}\fi
\expandafter\ifx\csname urlprefix\endcsname\relax\def\urlprefix{URL }\fi
\providecommand{\bibinfo}[2]{#2}
\providecommand{\eprint}[2][]{\url{#2}}

\bibitem[{\citenamefont{\mbox{R.~L.~Workman, {\it et al.}}}(2022)}]{pdg2022}
\bibinfo{author}{\bibnamefont{\mbox{R.~L.~Workman, {\it et al.}}}}
  (\bibinfo{collaboration}{Particle Data Group}), \bibinfo{journal}{PTEP}
  \textbf{\bibinfo{volume}{2022}}, \bibinfo{pages}{083C01}
  (\bibinfo{year}{2022}).

\bibitem[{\citenamefont{Bertone and Hooper}(2018)}]{dm_history}
\bibinfo{author}{\bibfnamefont{G.}~\bibnamefont{Bertone}} \bibnamefont{and}
  \bibinfo{author}{\bibfnamefont{D.}~\bibnamefont{Hooper}},
  \bibinfo{journal}{Rev. Mod. Phys.} \textbf{\bibinfo{volume}{90}},
  \bibinfo{pages}{045002} (\bibinfo{year}{2018}).

\bibitem[{\citenamefont{Jungman et~al.}(1996)\citenamefont{Jungman,
  Kamionkowski, and Griest}}]{susy_dm}
\bibinfo{author}{\bibfnamefont{G.}~\bibnamefont{Jungman}},
  \bibinfo{author}{\bibfnamefont{M.}~\bibnamefont{Kamionkowski}},
  \bibnamefont{and} \bibinfo{author}{\bibfnamefont{K.}~\bibnamefont{Griest}},
  \bibinfo{journal}{Phys. Rep.} \textbf{\bibinfo{volume}{267}},
  \bibinfo{pages}{195} (\bibinfo{year}{1996}).

\bibitem[{\citenamefont{Feng and Kumar}(2008)}]{wimpless}
\bibinfo{author}{\bibfnamefont{J.~L.} \bibnamefont{Feng}} \bibnamefont{and}
  \bibinfo{author}{\bibfnamefont{J.}~\bibnamefont{Kumar}},
  \bibinfo{journal}{Phys. Rev. Lett.} \textbf{\bibinfo{volume}{101}},
  \bibinfo{pages}{231301} (\bibinfo{year}{2008}).

\bibitem[{\citenamefont{Bernabei et~al.}(2008)}]{dama}
\bibinfo{author}{\bibfnamefont{R.}~\bibnamefont{Bernabei}} \bibnamefont{et~al.}
  (\bibinfo{collaboration}{DAMA}), \bibinfo{journal}{Eur. Phys. J. C}
  \textbf{\bibinfo{volume}{56}}, \bibinfo{pages}{333} (\bibinfo{year}{2008}).

\bibitem[{\citenamefont{Bernabei et~al.}(2010)}]{dama2}
\bibinfo{author}{\bibfnamefont{R.}~\bibnamefont{Bernabei}} \bibnamefont{et~al.}
  (\bibinfo{collaboration}{DAMA, LIBRA}), \bibinfo{journal}{Eur. Phys. J. C}
  \textbf{\bibinfo{volume}{67}}, \bibinfo{pages}{39} (\bibinfo{year}{2010}).

\bibitem[{\citenamefont{Aalseth et~al.}(2011)\citenamefont{Aalseth, Barbeau,
  Bowden, Cabrera-Palmer, Colaresi, Collar, Dazeley, de~Lurgio, Fast, Fields
  et~al.}}]{cogent}
\bibinfo{author}{\bibfnamefont{C.~E.} \bibnamefont{Aalseth}},
  \bibinfo{author}{\bibfnamefont{P.~S.} \bibnamefont{Barbeau}},
  \bibinfo{author}{\bibfnamefont{N.~S.} \bibnamefont{Bowden}},
  \bibinfo{author}{\bibfnamefont{B.}~\bibnamefont{Cabrera-Palmer}},
  \bibinfo{author}{\bibfnamefont{J.}~\bibnamefont{Colaresi}},
  \bibinfo{author}{\bibfnamefont{J.~I.} \bibnamefont{Collar}},
  \bibinfo{author}{\bibfnamefont{S.}~\bibnamefont{Dazeley}},
  \bibinfo{author}{\bibfnamefont{P.}~\bibnamefont{de~Lurgio}},
  \bibinfo{author}{\bibfnamefont{J.~E.} \bibnamefont{Fast}},
  \bibinfo{author}{\bibfnamefont{N.}~\bibnamefont{Fields}},
  \bibnamefont{et~al.} (\bibinfo{collaboration}{CoGeNT Collaboration}),
  \bibinfo{journal}{Phys. Rev. Lett.} \textbf{\bibinfo{volume}{106}},
  \bibinfo{pages}{131301} (\bibinfo{year}{2011}).

\bibitem[{\citenamefont{Aalseth et~al.}(2013)\citenamefont{Aalseth, Barbeau,
  Colaresi, Collar, Diaz~Leon, Fast, Fields, Hossbach, Knecht, Kos
  et~al.}}]{cogent2}
\bibinfo{author}{\bibfnamefont{C.~E.} \bibnamefont{Aalseth}},
  \bibinfo{author}{\bibfnamefont{P.~S.} \bibnamefont{Barbeau}},
  \bibinfo{author}{\bibfnamefont{J.}~\bibnamefont{Colaresi}},
  \bibinfo{author}{\bibfnamefont{J.~I.} \bibnamefont{Collar}},
  \bibinfo{author}{\bibfnamefont{J.}~\bibnamefont{Diaz~Leon}},
  \bibinfo{author}{\bibfnamefont{J.~E.} \bibnamefont{Fast}},
  \bibinfo{author}{\bibfnamefont{N.~E.} \bibnamefont{Fields}},
  \bibinfo{author}{\bibfnamefont{T.~W.} \bibnamefont{Hossbach}},
  \bibinfo{author}{\bibfnamefont{A.}~\bibnamefont{Knecht}},
  \bibinfo{author}{\bibfnamefont{M.~S.} \bibnamefont{Kos}},
  \bibnamefont{et~al.} (\bibinfo{collaboration}{CoGeNT Collaboration}),
  \bibinfo{journal}{Phys. Rev. D} \textbf{\bibinfo{volume}{88}},
  \bibinfo{pages}{012002} (\bibinfo{year}{2013}).

\bibitem[{\citenamefont{Agnese et~al.}(2013)\citenamefont{Agnese, Ahmed,
  Anderson, Arrenberg, Balakishiyeva, Basu~Thakur, Bauer, Billard, Borgland,
  Brandt et~al.}}]{cdms_si}
\bibinfo{author}{\bibfnamefont{R.}~\bibnamefont{Agnese}},
  \bibinfo{author}{\bibfnamefont{Z.}~\bibnamefont{Ahmed}},
  \bibinfo{author}{\bibfnamefont{A.~J.} \bibnamefont{Anderson}},
  \bibinfo{author}{\bibfnamefont{S.}~\bibnamefont{Arrenberg}},
  \bibinfo{author}{\bibfnamefont{D.}~\bibnamefont{Balakishiyeva}},
  \bibinfo{author}{\bibfnamefont{R.}~\bibnamefont{Basu~Thakur}},
  \bibinfo{author}{\bibfnamefont{D.~A.} \bibnamefont{Bauer}},
  \bibinfo{author}{\bibfnamefont{J.}~\bibnamefont{Billard}},
  \bibinfo{author}{\bibfnamefont{A.}~\bibnamefont{Borgland}},
  \bibinfo{author}{\bibfnamefont{D.}~\bibnamefont{Brandt}},
  \bibnamefont{et~al.} (\bibinfo{collaboration}{CDMS Collaboration}),
  \bibinfo{journal}{Phys. Rev. Lett.} \textbf{\bibinfo{volume}{111}},
  \bibinfo{pages}{251301} (\bibinfo{year}{2013}).

\bibitem[{\citenamefont{Akerib et~al.}(2017)\citenamefont{Akerib, Alsum,
  Ara\'ujo, Bai, Bailey, Balajthy, Beltrame, Bernard, Bernstein, Biesiadzinski
  et~al.}}]{lux}
\bibinfo{author}{\bibfnamefont{D.~S.} \bibnamefont{Akerib}},
  \bibinfo{author}{\bibfnamefont{S.}~\bibnamefont{Alsum}},
  \bibinfo{author}{\bibfnamefont{H.~M.} \bibnamefont{Ara\'ujo}},
  \bibinfo{author}{\bibfnamefont{X.}~\bibnamefont{Bai}},
  \bibinfo{author}{\bibfnamefont{A.~J.} \bibnamefont{Bailey}},
  \bibinfo{author}{\bibfnamefont{J.}~\bibnamefont{Balajthy}},
  \bibinfo{author}{\bibfnamefont{P.}~\bibnamefont{Beltrame}},
  \bibinfo{author}{\bibfnamefont{E.~P.} \bibnamefont{Bernard}},
  \bibinfo{author}{\bibfnamefont{A.}~\bibnamefont{Bernstein}},
  \bibinfo{author}{\bibfnamefont{T.~P.} \bibnamefont{Biesiadzinski}},
  \bibnamefont{et~al.} (\bibinfo{collaboration}{LUX Collaboration}),
  \bibinfo{journal}{Phys. Rev. Lett.} \textbf{\bibinfo{volume}{118}},
  \bibinfo{pages}{251302} (\bibinfo{year}{2017}).

\bibitem[{\citenamefont{Aprile et~al.}(2018)\citenamefont{Aprile, Aalbers,
  Agostini, Alfonsi, Althueser, Amaro, Anthony, Arneodo, Baudis, Bauermeister
  et~al.}}]{xenon_2018}
\bibinfo{author}{\bibfnamefont{E.}~\bibnamefont{Aprile}},
  \bibinfo{author}{\bibfnamefont{J.}~\bibnamefont{Aalbers}},
  \bibinfo{author}{\bibfnamefont{F.}~\bibnamefont{Agostini}},
  \bibinfo{author}{\bibfnamefont{M.}~\bibnamefont{Alfonsi}},
  \bibinfo{author}{\bibfnamefont{L.}~\bibnamefont{Althueser}},
  \bibinfo{author}{\bibfnamefont{F.~D.} \bibnamefont{Amaro}},
  \bibinfo{author}{\bibfnamefont{M.}~\bibnamefont{Anthony}},
  \bibinfo{author}{\bibfnamefont{F.}~\bibnamefont{Arneodo}},
  \bibinfo{author}{\bibfnamefont{L.}~\bibnamefont{Baudis}},
  \bibinfo{author}{\bibfnamefont{B.}~\bibnamefont{Bauermeister}},
  \bibnamefont{et~al.} (\bibinfo{collaboration}{XENON Collaboration 7}),
  \bibinfo{journal}{Phys. Rev. Lett.} \textbf{\bibinfo{volume}{121}},
  \bibinfo{pages}{111302} (\bibinfo{year}{2018}).

\bibitem[{\citenamefont{Meng et~al.}(2021)\citenamefont{Meng, Wang, Tao,
  Abdukerim, Bo, Chen, Chen, Chen, Cheng, Cheng et~al.}}]{pandax_4t}
\bibinfo{author}{\bibfnamefont{Y.}~\bibnamefont{Meng}},
  \bibinfo{author}{\bibfnamefont{Z.}~\bibnamefont{Wang}},
  \bibinfo{author}{\bibfnamefont{Y.}~\bibnamefont{Tao}},
  \bibinfo{author}{\bibfnamefont{A.}~\bibnamefont{Abdukerim}},
  \bibinfo{author}{\bibfnamefont{Z.}~\bibnamefont{Bo}},
  \bibinfo{author}{\bibfnamefont{W.}~\bibnamefont{Chen}},
  \bibinfo{author}{\bibfnamefont{X.}~\bibnamefont{Chen}},
  \bibinfo{author}{\bibfnamefont{Y.}~\bibnamefont{Chen}},
  \bibinfo{author}{\bibfnamefont{C.}~\bibnamefont{Cheng}},
  \bibinfo{author}{\bibfnamefont{Y.}~\bibnamefont{Cheng}}, \bibnamefont{et~al.}
  (\bibinfo{collaboration}{PandaX-4T Collaboration}), \bibinfo{journal}{Phys.
  Rev. Lett.} \textbf{\bibinfo{volume}{127}}, \bibinfo{pages}{261802}
  (\bibinfo{year}{2021}).

\bibitem[{\citenamefont{Peskin}(2015)}]{susy_lhc}
\bibinfo{author}{\bibfnamefont{M.~E.} \bibnamefont{Peskin}},
  \bibinfo{journal}{Proc.~Natl.~Acad.~Sci.~U.S.A.}
  \textbf{\bibinfo{volume}{112}}, \bibinfo{pages}{12256}
  (\bibinfo{year}{2015}).

\bibitem[{\citenamefont{Albert et~al.}(2017)\citenamefont{Albert, Anderson,
  Bechtol, Drlica-Wagner, Meyer, S{\'{a}}nchez-Conde, Strigari, Wood, Abbott,
  Abdalla et~al.}}]{fermiLAT}
\bibinfo{author}{\bibfnamefont{A.}~\bibnamefont{Albert}},
  \bibinfo{author}{\bibfnamefont{B.}~\bibnamefont{Anderson}},
  \bibinfo{author}{\bibfnamefont{K.}~\bibnamefont{Bechtol}},
  \bibinfo{author}{\bibfnamefont{A.}~\bibnamefont{Drlica-Wagner}},
  \bibinfo{author}{\bibfnamefont{M.}~\bibnamefont{Meyer}},
  \bibinfo{author}{\bibfnamefont{M.}~\bibnamefont{S{\'{a}}nchez-Conde}},
  \bibinfo{author}{\bibfnamefont{L.}~\bibnamefont{Strigari}},
  \bibinfo{author}{\bibfnamefont{M.}~\bibnamefont{Wood}},
  \bibinfo{author}{\bibfnamefont{T.~M.~C.} \bibnamefont{Abbott}},
  \bibinfo{author}{\bibfnamefont{F.~B.} \bibnamefont{Abdalla}},
  \bibnamefont{et~al.}, \bibinfo{journal}{Astrophys.~J.~}
  \textbf{\bibinfo{volume}{834}}, \bibinfo{pages}{110} (\bibinfo{year}{2017}).

\bibitem[{\citenamefont{Hooper and Goodenough}(2011)}]{hooper}
\bibinfo{author}{\bibfnamefont{D.}~\bibnamefont{Hooper}} \bibnamefont{and}
  \bibinfo{author}{\bibfnamefont{L.}~\bibnamefont{Goodenough}},
  \bibinfo{journal}{Phys.~Lett.~ B} \textbf{\bibinfo{volume}{697}},
  \bibinfo{pages}{412} (\bibinfo{year}{2011}).

\bibitem[{\citenamefont{Aguilar et~al.}(2016)\citenamefont{Aguilar,
  Ali~Cavasonza, Alpat, Ambrosi, Arruda, Attig, Aupetit, Azzarello,
  Bachlechner, Barao et~al.}}]{ams02}
\bibinfo{author}{\bibfnamefont{M.}~\bibnamefont{Aguilar}},
  \bibinfo{author}{\bibfnamefont{L.}~\bibnamefont{Ali~Cavasonza}},
  \bibinfo{author}{\bibfnamefont{B.}~\bibnamefont{Alpat}},
  \bibinfo{author}{\bibfnamefont{G.}~\bibnamefont{Ambrosi}},
  \bibinfo{author}{\bibfnamefont{L.}~\bibnamefont{Arruda}},
  \bibinfo{author}{\bibfnamefont{N.}~\bibnamefont{Attig}},
  \bibinfo{author}{\bibfnamefont{S.}~\bibnamefont{Aupetit}},
  \bibinfo{author}{\bibfnamefont{P.}~\bibnamefont{Azzarello}},
  \bibinfo{author}{\bibfnamefont{A.}~\bibnamefont{Bachlechner}},
  \bibinfo{author}{\bibfnamefont{F.}~\bibnamefont{Barao}}, \bibnamefont{et~al.}
  (\bibinfo{collaboration}{AMS Collaboration}), \bibinfo{journal}{Phys. Rev.
  Lett.} \textbf{\bibinfo{volume}{117}}, \bibinfo{pages}{091103}
  (\bibinfo{year}{2016}).

\bibitem[{\citenamefont{Cuoco et~al.}(2017)\citenamefont{Cuoco, Kr\"amer, and
  Korsmeier}}]{cuoco}
\bibinfo{author}{\bibfnamefont{A.}~\bibnamefont{Cuoco}},
  \bibinfo{author}{\bibfnamefont{M.}~\bibnamefont{Kr\"amer}}, \bibnamefont{and}
  \bibinfo{author}{\bibfnamefont{M.}~\bibnamefont{Korsmeier}},
  \bibinfo{journal}{Phys. Rev. Lett.} \textbf{\bibinfo{volume}{118}},
  \bibinfo{pages}{191102} (\bibinfo{year}{2017}).

\bibitem[{\citenamefont{Marsh}(2016)}]{axions}
\bibinfo{author}{\bibfnamefont{D.~J.} \bibnamefont{Marsh}},
  \bibinfo{journal}{Phys.~Rep.~} \textbf{\bibinfo{volume}{643}},
  \bibinfo{pages}{1} (\bibinfo{year}{2016}).

\bibitem[{\citenamefont{Roszkowski et~al.}(2018)\citenamefont{Roszkowski,
  Sessolo, and Trojanowski}}]{roszkowski}
\bibinfo{author}{\bibfnamefont{L.}~\bibnamefont{Roszkowski}},
  \bibinfo{author}{\bibfnamefont{E.~M.} \bibnamefont{Sessolo}},
  \bibnamefont{and}
  \bibinfo{author}{\bibfnamefont{S.}~\bibnamefont{Trojanowski}},
  \bibinfo{journal}{Rept. Prog. Phys.} \textbf{\bibinfo{volume}{81}},
  \bibinfo{pages}{066201} (\bibinfo{year}{2018}).

\bibitem[{\citenamefont{Arcadi et~al.}(2018)\citenamefont{Arcadi, Dutra, Ghosh,
  Lindner, Mambrini, Pierre, Profumo, and Queiroz}}]{arcadi}
\bibinfo{author}{\bibfnamefont{G.}~\bibnamefont{Arcadi}},
  \bibinfo{author}{\bibfnamefont{M.}~\bibnamefont{Dutra}},
  \bibinfo{author}{\bibfnamefont{P.}~\bibnamefont{Ghosh}},
  \bibinfo{author}{\bibfnamefont{M.}~\bibnamefont{Lindner}},
  \bibinfo{author}{\bibfnamefont{Y.}~\bibnamefont{Mambrini}},
  \bibinfo{author}{\bibfnamefont{M.}~\bibnamefont{Pierre}},
  \bibinfo{author}{\bibfnamefont{S.}~\bibnamefont{Profumo}}, \bibnamefont{and}
  \bibinfo{author}{\bibfnamefont{F.~S.} \bibnamefont{Queiroz}},
  \bibinfo{journal}{Eur. Phys. J. C} \textbf{\bibinfo{volume}{78}},
  \bibinfo{pages}{203} (\bibinfo{year}{2018}).

\bibitem[{\citenamefont{Blanco et~al.}(2019)\citenamefont{Blanco, Escudero,
  Hooper, and Witte}}]{blanco}
\bibinfo{author}{\bibfnamefont{C.}~\bibnamefont{Blanco}},
  \bibinfo{author}{\bibfnamefont{M.}~\bibnamefont{Escudero}},
  \bibinfo{author}{\bibfnamefont{D.}~\bibnamefont{Hooper}}, \bibnamefont{and}
  \bibinfo{author}{\bibfnamefont{S.~J.} \bibnamefont{Witte}},
  \bibinfo{journal}{Journal of Cosmology and Astroparticle Physics}
  \textbf{\bibinfo{volume}{2019}}, \bibinfo{pages}{024} (\bibinfo{year}{2019}).

\bibitem[{\citenamefont{De~Simone and Jacques}(2016)}]{desimone}
\bibinfo{author}{\bibfnamefont{A.}~\bibnamefont{De~Simone}} \bibnamefont{and}
  \bibinfo{author}{\bibfnamefont{T.}~\bibnamefont{Jacques}},
  \bibinfo{journal}{Eur. Phys. J.} \textbf{\bibinfo{volume}{C76}},
  \bibinfo{pages}{367} (\bibinfo{year}{2016}).

\bibitem[{\citenamefont{Raby and West}(1987)}]{raby1987}
\bibinfo{author}{\bibfnamefont{S.}~\bibnamefont{Raby}} \bibnamefont{and}
  \bibinfo{author}{\bibfnamefont{G.~B.} \bibnamefont{West}},
  \bibinfo{journal}{Phys.~Lett.~ B} \textbf{\bibinfo{volume}{194}},
  \bibinfo{pages}{557} (\bibinfo{year}{1987}).

\bibitem[{\citenamefont{Raby and West}(1988)}]{raby1988}
\bibinfo{author}{\bibfnamefont{S.}~\bibnamefont{Raby}} \bibnamefont{and}
  \bibinfo{author}{\bibfnamefont{G.}~\bibnamefont{West}},
  \bibinfo{journal}{Phys.~Lett.~ B} \textbf{\bibinfo{volume}{200}},
  \bibinfo{pages}{547} (\bibinfo{year}{1988}).

\bibitem[{\citenamefont{Pospelov and ter Veldhuis}(2000)}]{pospelov}
\bibinfo{author}{\bibfnamefont{M.}~\bibnamefont{Pospelov}} \bibnamefont{and}
  \bibinfo{author}{\bibfnamefont{T.}~\bibnamefont{ter Veldhuis}},
  \bibinfo{journal}{Phys. Lett.} \textbf{\bibinfo{volume}{B480}},
  \bibinfo{pages}{181} (\bibinfo{year}{2000}).

\bibitem[{\citenamefont{Sigurdson et~al.}(2004)\citenamefont{Sigurdson, Doran,
  Kurylov, Caldwell, and Kamionkowski}}]{sigurdson2004}
\bibinfo{author}{\bibfnamefont{K.}~\bibnamefont{Sigurdson}},
  \bibinfo{author}{\bibfnamefont{M.}~\bibnamefont{Doran}},
  \bibinfo{author}{\bibfnamefont{A.}~\bibnamefont{Kurylov}},
  \bibinfo{author}{\bibfnamefont{R.~R.} \bibnamefont{Caldwell}},
  \bibnamefont{and}
  \bibinfo{author}{\bibfnamefont{M.}~\bibnamefont{Kamionkowski}},
  \bibinfo{journal}{Phys. Rev. D} \textbf{\bibinfo{volume}{70}},
  \bibinfo{pages}{083501} (\bibinfo{year}{2004}).

\bibitem[{\citenamefont{Sigurdson et~al.}(2006)\citenamefont{Sigurdson, Doran,
  Kurylov, Caldwell, and Kamionkowski}}]{sigurdson2004err}
\bibinfo{author}{\bibfnamefont{K.}~\bibnamefont{Sigurdson}},
  \bibinfo{author}{\bibfnamefont{M.}~\bibnamefont{Doran}},
  \bibinfo{author}{\bibfnamefont{A.}~\bibnamefont{Kurylov}},
  \bibinfo{author}{\bibfnamefont{R.~R.} \bibnamefont{Caldwell}},
  \bibnamefont{and}
  \bibinfo{author}{\bibfnamefont{M.}~\bibnamefont{Kamionkowski}},
  \bibinfo{journal}{Phys. Rev. D} \textbf{\bibinfo{volume}{73}},
  \bibinfo{pages}{089903} (\bibinfo{year}{2006}).

\bibitem[{\citenamefont{Mass\'o et~al.}(2009)\citenamefont{Mass\'o, Mohanty,
  and Rao}}]{masso2009}
\bibinfo{author}{\bibfnamefont{E.}~\bibnamefont{Mass\'o}},
  \bibinfo{author}{\bibfnamefont{S.}~\bibnamefont{Mohanty}}, \bibnamefont{and}
  \bibinfo{author}{\bibfnamefont{S.}~\bibnamefont{Rao}},
  \bibinfo{journal}{Phys. Rev. D} \textbf{\bibinfo{volume}{80}},
  \bibinfo{pages}{036009} (\bibinfo{year}{2009}).

\bibitem[{\citenamefont{Heo}(2010)}]{heo2010}
\bibinfo{author}{\bibfnamefont{J.~H.} \bibnamefont{Heo}},
  \bibinfo{journal}{Phys.~Lett.~ B} \textbf{\bibinfo{volume}{693}},
  \bibinfo{pages}{255} (\bibinfo{year}{2010}).

\bibitem[{\citenamefont{Fitzpatrick and Zurek}(2010)}]{fitzpatrick2010}
\bibinfo{author}{\bibfnamefont{A.~L.} \bibnamefont{Fitzpatrick}}
  \bibnamefont{and} \bibinfo{author}{\bibfnamefont{K.~M.} \bibnamefont{Zurek}},
  \bibinfo{journal}{Phys. Rev. D} \textbf{\bibinfo{volume}{82}},
  \bibinfo{pages}{075004} (\bibinfo{year}{2010}).

\bibitem[{\citenamefont{Barger et~al.}(2011)\citenamefont{Barger, Keung, and
  Marfatia}}]{barger2011}
\bibinfo{author}{\bibfnamefont{V.}~\bibnamefont{Barger}},
  \bibinfo{author}{\bibfnamefont{W.-Y.} \bibnamefont{Keung}}, \bibnamefont{and}
  \bibinfo{author}{\bibfnamefont{D.}~\bibnamefont{Marfatia}},
  \bibinfo{journal}{Phys.~Lett.~ B} \textbf{\bibinfo{volume}{696}},
  \bibinfo{pages}{74} (\bibinfo{year}{2011}).

\bibitem[{\citenamefont{Barger et~al.}(2012)\citenamefont{Barger, Keung,
  Marfatia, and Tseng}}]{barger2012}
\bibinfo{author}{\bibfnamefont{V.}~\bibnamefont{Barger}},
  \bibinfo{author}{\bibfnamefont{W.-Y.} \bibnamefont{Keung}},
  \bibinfo{author}{\bibfnamefont{D.}~\bibnamefont{Marfatia}}, \bibnamefont{and}
  \bibinfo{author}{\bibfnamefont{P.-Y.} \bibnamefont{Tseng}},
  \bibinfo{journal}{Phys.~Lett.~ B} \textbf{\bibinfo{volume}{717}},
  \bibinfo{pages}{219} (\bibinfo{year}{2012}).

\bibitem[{\citenamefont{Fortin and Tait}(2012)}]{fortin2012}
\bibinfo{author}{\bibfnamefont{J.-F. m.~c.} \bibnamefont{Fortin}}
  \bibnamefont{and} \bibinfo{author}{\bibfnamefont{T.~M.~P.}
  \bibnamefont{Tait}}, \bibinfo{journal}{Phys. Rev. D}
  \textbf{\bibinfo{volume}{85}}, \bibinfo{pages}{063506}
  (\bibinfo{year}{2012}).

\bibitem[{\citenamefont{Weiner and Yavin}(2013)}]{weiner2013}
\bibinfo{author}{\bibfnamefont{N.}~\bibnamefont{Weiner}} \bibnamefont{and}
  \bibinfo{author}{\bibfnamefont{I.}~\bibnamefont{Yavin}},
  \bibinfo{journal}{Phys. Rev. D} \textbf{\bibinfo{volume}{87}},
  \bibinfo{pages}{023523} (\bibinfo{year}{2013}).

\bibitem[{\citenamefont{Del~Nobile et~al.}(2014)\citenamefont{Del~Nobile,
  Gelmini, Gondolo, and Huh}}]{delnobile2014}
\bibinfo{author}{\bibfnamefont{E.}~\bibnamefont{Del~Nobile}},
  \bibinfo{author}{\bibfnamefont{G.~B.} \bibnamefont{Gelmini}},
  \bibinfo{author}{\bibfnamefont{P.}~\bibnamefont{Gondolo}}, \bibnamefont{and}
  \bibinfo{author}{\bibfnamefont{J.-H.} \bibnamefont{Huh}},
  \bibinfo{journal}{JCAP} \textbf{\bibinfo{volume}{1406}}, \bibinfo{pages}{002}
  (\bibinfo{year}{2014}).

\bibitem[{\citenamefont{Kopp et~al.}(2014)\citenamefont{Kopp, Michaels, and
  Smirnov}}]{kopp2014}
\bibinfo{author}{\bibfnamefont{J.}~\bibnamefont{Kopp}},
  \bibinfo{author}{\bibfnamefont{L.}~\bibnamefont{Michaels}}, \bibnamefont{and}
  \bibinfo{author}{\bibfnamefont{J.}~\bibnamefont{Smirnov}},
  \bibinfo{journal}{JCAP} \textbf{\bibinfo{volume}{04}}, \bibinfo{pages}{022}
  (\bibinfo{year}{2014}).

\bibitem[{\citenamefont{Gresham and Zurek}(2014)}]{gresham2014}
\bibinfo{author}{\bibfnamefont{M.~I.} \bibnamefont{Gresham}} \bibnamefont{and}
  \bibinfo{author}{\bibfnamefont{K.~M.} \bibnamefont{Zurek}},
  \bibinfo{journal}{Phys. Rev. D} \textbf{\bibinfo{volume}{89}},
  \bibinfo{pages}{016017} (\bibinfo{year}{2014}).

\bibitem[{\citenamefont{Matsumoto et~al.}(2014)\citenamefont{Matsumoto,
  Mukhopadhyay, and Tsai}}]{matsumoto2014}
\bibinfo{author}{\bibfnamefont{S.}~\bibnamefont{Matsumoto}},
  \bibinfo{author}{\bibfnamefont{S.}~\bibnamefont{Mukhopadhyay}},
  \bibnamefont{and} \bibinfo{author}{\bibfnamefont{Y.-L.~S.}
  \bibnamefont{Tsai}}, \bibinfo{journal}{JHEP} \textbf{\bibinfo{volume}{10}},
  \bibinfo{pages}{155} (\bibinfo{year}{2014}).

\bibitem[{\citenamefont{Antipin et~al.}(2015)\citenamefont{Antipin, Redi,
  Strumia, and Vigiani}}]{antipin2015}
\bibinfo{author}{\bibfnamefont{O.}~\bibnamefont{Antipin}},
  \bibinfo{author}{\bibfnamefont{M.}~\bibnamefont{Redi}},
  \bibinfo{author}{\bibfnamefont{A.}~\bibnamefont{Strumia}}, \bibnamefont{and}
  \bibinfo{author}{\bibfnamefont{E.}~\bibnamefont{Vigiani}},
  \bibinfo{journal}{JHEP} \textbf{\bibinfo{volume}{07}}, \bibinfo{pages}{039}
  (\bibinfo{year}{2015}).

\bibitem[{\citenamefont{Kayser}(1984)}]{bk84}
\bibinfo{author}{\bibfnamefont{B.}~\bibnamefont{Kayser}},
  \bibinfo{journal}{Phys. Rev.} \textbf{\bibinfo{volume}{D30}},
  \bibinfo{pages}{1023} (\bibinfo{year}{1984}).

\bibitem[{\citenamefont{Nieves}(1982)}]{nieves}
\bibinfo{author}{\bibfnamefont{J.~F.} \bibnamefont{Nieves}},
  \bibinfo{journal}{Phys. Rev.} \textbf{\bibinfo{volume}{D26}},
  \bibinfo{pages}{3152} (\bibinfo{year}{1982}).

\bibitem[{\citenamefont{Ho and Scherrer}(2013)}]{ho2013}
\bibinfo{author}{\bibfnamefont{C.~M.} \bibnamefont{Ho}} \bibnamefont{and}
  \bibinfo{author}{\bibfnamefont{R.~J.} \bibnamefont{Scherrer}},
  \bibinfo{journal}{Phys.~Lett.~ B} \textbf{\bibinfo{volume}{722}},
  \bibinfo{pages}{341} (\bibinfo{year}{2013}).

\bibitem[{\citenamefont{Gao et~al.}(2014)\citenamefont{Gao, Ho, and
  Scherrer}}]{gao2014}
\bibinfo{author}{\bibfnamefont{Y.}~\bibnamefont{Gao}},
  \bibinfo{author}{\bibfnamefont{C.~M.} \bibnamefont{Ho}}, \bibnamefont{and}
  \bibinfo{author}{\bibfnamefont{R.~J.} \bibnamefont{Scherrer}},
  \bibinfo{journal}{Phys. Rev. D} \textbf{\bibinfo{volume}{89}},
  \bibinfo{pages}{045006} (\bibinfo{year}{2014}).

\bibitem[{\citenamefont{Sandick et~al.}(2016)\citenamefont{Sandick, Sinha, and
  Teng}}]{sandick2016}
\bibinfo{author}{\bibfnamefont{P.}~\bibnamefont{Sandick}},
  \bibinfo{author}{\bibfnamefont{K.}~\bibnamefont{Sinha}}, \bibnamefont{and}
  \bibinfo{author}{\bibfnamefont{F.}~\bibnamefont{Teng}},
  \bibinfo{journal}{JHEP} \textbf{\bibinfo{volume}{10}}, \bibinfo{pages}{018}
  (\bibinfo{year}{2016}).

\bibitem[{\citenamefont{Gelmini}(2017)}]{gelmini2017}
\bibinfo{author}{\bibfnamefont{G.~B.} \bibnamefont{Gelmini}},
  \bibinfo{journal}{Rep.~Prog.~Phys.~} \textbf{\bibinfo{volume}{80}},
  \bibinfo{pages}{082201} (\bibinfo{year}{2017}).

\bibitem[{\citenamefont{Alves et~al.}(2018)\citenamefont{Alves, Santos, and
  Sinha}}]{alves}
\bibinfo{author}{\bibfnamefont{A.}~\bibnamefont{Alves}},
  \bibinfo{author}{\bibfnamefont{A.~C.~O.} \bibnamefont{Santos}},
  \bibnamefont{and} \bibinfo{author}{\bibfnamefont{K.}~\bibnamefont{Sinha}},
  \bibinfo{journal}{Phys. Rev. D} \textbf{\bibinfo{volume}{97}},
  \bibinfo{pages}{055023} (\bibinfo{year}{2018}).

\bibitem[{\citenamefont{Ragusa}(1993)}]{ragusa}
\bibinfo{author}{\bibfnamefont{S.}~\bibnamefont{Ragusa}},
  \bibinfo{journal}{Phys. Rev.} \textbf{\bibinfo{volume}{D47}},
  \bibinfo{pages}{3757} (\bibinfo{year}{1993}).

\bibitem[{\citenamefont{Gorchtein}(2008)}]{gorchtein}
\bibinfo{author}{\bibfnamefont{M.}~\bibnamefont{Gorchtein}},
  \bibinfo{journal}{Phys. Rev.} \textbf{\bibinfo{volume}{C77}},
  \bibinfo{pages}{065501} (\bibinfo{year}{2008}).

\bibitem[{\citenamefont{Latimer}(2016)}]{maj_2photon}
\bibinfo{author}{\bibfnamefont{D.~C.} \bibnamefont{Latimer}},
  \bibinfo{journal}{Phys. Rev.} \textbf{\bibinfo{volume}{D94}},
  \bibinfo{pages}{093010} (\bibinfo{year}{2016}).

\bibitem[{\citenamefont{Latimer}(2017)}]{anapole_2photon}
\bibinfo{author}{\bibfnamefont{D.~C.} \bibnamefont{Latimer}},
  \bibinfo{journal}{Phys. Rev. D} \textbf{\bibinfo{volume}{95}},
  \bibinfo{pages}{095023} (\bibinfo{year}{2017}).

\bibitem[{\citenamefont{Walter et~al.}(2022)\citenamefont{Walter, Hall, and
  Latimer}}]{walter}
\bibinfo{author}{\bibfnamefont{K.}~\bibnamefont{Walter}},
  \bibinfo{author}{\bibfnamefont{K.}~\bibnamefont{Hall}}, \bibnamefont{and}
  \bibinfo{author}{\bibfnamefont{D.~C.} \bibnamefont{Latimer}},
  \bibinfo{journal}{Phys. Rev. D} \textbf{\bibinfo{volume}{106}},
  \bibinfo{pages}{096021} (\bibinfo{year}{2022}).

\bibitem[{\citenamefont{Zeldovich}(1957)}]{zeldovich}
\bibinfo{author}{\bibfnamefont{{\relax Ya}.~B.} \bibnamefont{Zeldovich}},
  \bibinfo{journal}{Sov. Phys. JETP} \textbf{\bibinfo{volume}{6}},
  \bibinfo{pages}{1184} (\bibinfo{year}{1957}), \bibinfo{note}{[Zh. Eksp. Teor.
  Fiz. {\bf 33},1531 (1957)]}.

\bibitem[{\citenamefont{Giunti and Studenikin}(2015)}]{neutrino_EM_review}
\bibinfo{author}{\bibfnamefont{C.}~\bibnamefont{Giunti}} \bibnamefont{and}
  \bibinfo{author}{\bibfnamefont{A.}~\bibnamefont{Studenikin}},
  \bibinfo{journal}{Rev. Mod. Phys.} \textbf{\bibinfo{volume}{87}},
  \bibinfo{pages}{531} (\bibinfo{year}{2015}).

\bibitem[{\citenamefont{Cabral-Rosetti
  et~al.}(2016)\citenamefont{Cabral-Rosetti, Mondrag\'on, and
  Reyes-P\'erez}}]{nulino_anapole}
\bibinfo{author}{\bibfnamefont{L.~G.} \bibnamefont{Cabral-Rosetti}},
  \bibinfo{author}{\bibfnamefont{M.}~\bibnamefont{Mondrag\'on}},
  \bibnamefont{and}
  \bibinfo{author}{\bibfnamefont{E.}~\bibnamefont{Reyes-P\'erez}},
  \bibinfo{journal}{Nucl. Phys.} \textbf{\bibinfo{volume}{B907}},
  \bibinfo{pages}{1} (\bibinfo{year}{2016}).

\bibitem[{\citenamefont{Prange}(1958)}]{prange}
\bibinfo{author}{\bibfnamefont{R.~E.} \bibnamefont{Prange}},
  \bibinfo{journal}{Phys. Rev.} \textbf{\bibinfo{volume}{110}},
  \bibinfo{pages}{240} (\bibinfo{year}{1958}).

\bibitem[{\citenamefont{Kolb and Turner}(1990)}]{kolb_turner}
\bibinfo{author}{\bibfnamefont{E.}~\bibnamefont{Kolb}} \bibnamefont{and}
  \bibinfo{author}{\bibfnamefont{M.}~\bibnamefont{Turner}},
  \emph{\bibinfo{title}{The Early Universe}}
  (\bibinfo{publisher}{Addison-Wesley}, \bibinfo{address}{Redwood City, CA},
  \bibinfo{year}{1990}).

\bibitem[{\citenamefont{Scherrer and Turner}(1986)}]{Scherrer:1985zt}
\bibinfo{author}{\bibfnamefont{R.~J.} \bibnamefont{Scherrer}} \bibnamefont{and}
  \bibinfo{author}{\bibfnamefont{M.~S.} \bibnamefont{Turner}},
  \bibinfo{journal}{Phys. Rev.} \textbf{\bibinfo{volume}{D33}},
  \bibinfo{pages}{1585} (\bibinfo{year}{1986}), \bibinfo{note}{[Erratum: Phys.
  Rev. {\bf D34}, 3263 (1986)]}.

\bibitem[{\citenamefont{Ackermann et~al.}(2015)\citenamefont{Ackermann, Ajello,
  Albert, Anderson, Atwood, Baldini, Barbiellini, Bastieri, Bellazzini,
  Bissaldi et~al.}}]{ackermann}
\bibinfo{author}{\bibfnamefont{M.}~\bibnamefont{Ackermann}},
  \bibinfo{author}{\bibfnamefont{M.}~\bibnamefont{Ajello}},
  \bibinfo{author}{\bibfnamefont{A.}~\bibnamefont{Albert}},
  \bibinfo{author}{\bibfnamefont{B.}~\bibnamefont{Anderson}},
  \bibinfo{author}{\bibfnamefont{W.~B.} \bibnamefont{Atwood}},
  \bibinfo{author}{\bibfnamefont{L.}~\bibnamefont{Baldini}},
  \bibinfo{author}{\bibfnamefont{G.}~\bibnamefont{Barbiellini}},
  \bibinfo{author}{\bibfnamefont{D.}~\bibnamefont{Bastieri}},
  \bibinfo{author}{\bibfnamefont{R.}~\bibnamefont{Bellazzini}},
  \bibinfo{author}{\bibfnamefont{E.}~\bibnamefont{Bissaldi}},
  \bibnamefont{et~al.}, \bibinfo{journal}{Phys. Rev. D}
  \textbf{\bibinfo{volume}{91}}, \bibinfo{pages}{122002}
  (\bibinfo{year}{2015}).

\bibitem[{\citenamefont{D'Amico et~al.}(2018)\citenamefont{D'Amico, Panci, and
  Strumia}}]{damico}
\bibinfo{author}{\bibfnamefont{G.}~\bibnamefont{D'Amico}},
  \bibinfo{author}{\bibfnamefont{P.}~\bibnamefont{Panci}}, \bibnamefont{and}
  \bibinfo{author}{\bibfnamefont{A.}~\bibnamefont{Strumia}},
  \bibinfo{journal}{Phys. Rev. Lett.} \textbf{\bibinfo{volume}{121}},
  \bibinfo{pages}{011103} (\bibinfo{year}{2018}).

\bibitem[{\citenamefont{Slatyer}(2016)}]{slatyer}
\bibinfo{author}{\bibfnamefont{T.~R.} \bibnamefont{Slatyer}},
  \bibinfo{journal}{Phys. Rev. D} \textbf{\bibinfo{volume}{93}},
  \bibinfo{pages}{023527} (\bibinfo{year}{2016}).

\bibitem[{\citenamefont{Ade et~al.}(2016)}]{planck2015_cosmo}
\bibinfo{author}{\bibfnamefont{P.~A.~R.} \bibnamefont{Ade}}
  \bibnamefont{et~al.} (\bibinfo{collaboration}{Planck}),
  \bibinfo{journal}{Astron. Astrophys.} \textbf{\bibinfo{volume}{594}},
  \bibinfo{pages}{A13} (\bibinfo{year}{2016}).

\bibitem[{\citenamefont{Steigman et~al.}(2012)\citenamefont{Steigman, Dasgupta,
  and Beacom}}]{steigman2012}
\bibinfo{author}{\bibfnamefont{G.}~\bibnamefont{Steigman}},
  \bibinfo{author}{\bibfnamefont{B.}~\bibnamefont{Dasgupta}}, \bibnamefont{and}
  \bibinfo{author}{\bibfnamefont{J.~F.} \bibnamefont{Beacom}},
  \bibinfo{journal}{Phys. Rev. D} \textbf{\bibinfo{volume}{86}},
  \bibinfo{pages}{023506} (\bibinfo{year}{2012}).

\bibitem[{\citenamefont{Turner and Widrow}(1988)}]{PhysRevD.37.2743}
\bibinfo{author}{\bibfnamefont{M.~S.} \bibnamefont{Turner}} \bibnamefont{and}
  \bibinfo{author}{\bibfnamefont{L.~M.} \bibnamefont{Widrow}},
  \bibinfo{journal}{Phys. Rev. D} \textbf{\bibinfo{volume}{37}},
  \bibinfo{pages}{2743} (\bibinfo{year}{1988}).

\bibitem[{\citenamefont{Sigl et~al.}(1997)\citenamefont{Sigl, Olinto, and
  Jedamzik}}]{PhysRevD.55.4582}
\bibinfo{author}{\bibfnamefont{G.}~\bibnamefont{Sigl}},
  \bibinfo{author}{\bibfnamefont{A.~V.} \bibnamefont{Olinto}},
  \bibnamefont{and} \bibinfo{author}{\bibfnamefont{K.}~\bibnamefont{Jedamzik}},
  \bibinfo{journal}{Phys. Rev. D} \textbf{\bibinfo{volume}{55}},
  \bibinfo{pages}{4582} (\bibinfo{year}{1997}).

\bibitem[{\citenamefont{Baym et~al.}(1996)\citenamefont{Baym, B\"odeker, and
  McLerran}}]{PhysRevD.53.662}
\bibinfo{author}{\bibfnamefont{G.}~\bibnamefont{Baym}},
  \bibinfo{author}{\bibfnamefont{D.}~\bibnamefont{B\"odeker}},
  \bibnamefont{and} \bibinfo{author}{\bibfnamefont{L.}~\bibnamefont{McLerran}},
  \bibinfo{journal}{Phys. Rev. D} \textbf{\bibinfo{volume}{53}},
  \bibinfo{pages}{662} (\bibinfo{year}{1996}).

\bibitem[{\citenamefont{{Quashnock} et~al.}(1989)\citenamefont{{Quashnock},
  {Loeb}, and {Spergel}}}]{QCDtrans_currents}
\bibinfo{author}{\bibfnamefont{J.~M.} \bibnamefont{{Quashnock}}},
  \bibinfo{author}{\bibfnamefont{A.}~\bibnamefont{{Loeb}}}, \bibnamefont{and}
  \bibinfo{author}{\bibfnamefont{D.~N.} \bibnamefont{{Spergel}}},
  \bibinfo{journal}{Astrophys.~J.~Lett.~} \textbf{\bibinfo{volume}{344}},
  \bibinfo{pages}{L49} (\bibinfo{year}{1989}).

\bibitem[{\citenamefont{Cheng and Olinto}(1994)}]{PhysRevD.50.2421}
\bibinfo{author}{\bibfnamefont{B.}~\bibnamefont{Cheng}} \bibnamefont{and}
  \bibinfo{author}{\bibfnamefont{A.~V.} \bibnamefont{Olinto}},
  \bibinfo{journal}{Phys. Rev. D} \textbf{\bibinfo{volume}{50}},
  \bibinfo{pages}{2421} (\bibinfo{year}{1994}).

\bibitem[{\citenamefont{de~Souza and Opher}(2008)}]{PhysRevD.77.043529}
\bibinfo{author}{\bibfnamefont{R.~S.} \bibnamefont{de~Souza}} \bibnamefont{and}
  \bibinfo{author}{\bibfnamefont{R.}~\bibnamefont{Opher}},
  \bibinfo{journal}{Phys. Rev. D} \textbf{\bibinfo{volume}{77}},
  \bibinfo{pages}{043529} (\bibinfo{year}{2008}).

\bibitem[{\citenamefont{Siegel and Fry}(2006)}]{siegel_fry}
\bibinfo{author}{\bibfnamefont{E.~R.} \bibnamefont{Siegel}} \bibnamefont{and}
  \bibinfo{author}{\bibfnamefont{J.~N.} \bibnamefont{Fry}}
  (\bibinfo{year}{2006}),
  \urlprefix\url{https://arxiv.org/abs/astro-ph/0609031}.

\end{thebibliography}

\end{document}